\documentclass[11pt]{article}
\usepackage{macros}

\usepackage[citestyle=alphabetic,bibstyle=alphabetic,minalphanames=5,maxalphanames=6, maxbibnames=99,url=false]{biblatex}
\DeclareLabelalphaTemplate{
  \labelelement{
    \field[final]{shorthand}
    \field{label}
    \field[strwidth=3,strside=left,ifnames=1]{labelname}
    \field[strwidth=1,strside=left]{labelname}
  }
  \labelelement{
    \field[strwidth=2,strside=right]{origyear}
    \field[strwidth=2,strside=right]{year}
  }
}
\providecommand{\thresholddim}{}
\renewcommand{\thresholddim}{k^*}
\providecommand{\expectationdim}{}
\renewcommand{\expectationdim}{k^{\mathsf{exp}}}
\providecommand{\projectionexpectationdim}{}
\renewcommand{\projectionexpectationdim}{\expectationdim_{\mathrm{proj}}}

\providecommand{\covercost}{}
\renewcommand{\covercost}{\operatorname{\mathsf{cost}}}
\providecommand{\covers}{}
\renewcommand{\covers}{\operatorname{\mathsf{covers}}}

\providecommand{\pseudorandom}[2]{}
\renewcommand{\pseudorandom}[2]{\ensuremath{\inparen{#1,#2}}\ifmmode\text{-pseudorandom}\else-pseudorandom\fi}
\providecommand{\almostpseudorandom}[3]{}
\renewcommand{\almostpseudorandom}[3]{\ensuremath{\inparen{#1,#2,#3}}\ifmmode\text{-conditionally-pseudorandom}\else-conditionally-pseudorandom\fi}

\title{A Universal Threshold Theory for \\Local Properties of Random Linear Codes}

\date{}
\author{Vinayak M. Kumar\thanks{
Simons Institute for the Theory of Computing.
\href{mailto:vmkumar@berkeley.edu}{\texttt{vmkumar@berkeley.edu}}.
Supported by a Simons Institute Research Fellowship.
Part of this work was conducted while the author was a graduate student at UT Austin, supported by a Jane Street Fellowship and a UT Austin Dean's Prestigious Fellowship
Supplement.}
  \qquad Ray Li\thanks{Santa Clara University. \href{mailto:rayyli@alumni.stanford.edu}{\texttt{rayyli@alumni.stanford.edu}}. Supported by National Science Foundation grant CCF-2347371.
  }
  \qquad Jonathan Mosheiff\thanks{Ben-Gurion University of the Negev. \href{mailto:mosheiff@bgu.ac.il}{\texttt{mosheiff@bgu.ac.il}}. Supported by Israel Science Foundation grant \#3450/24.
}
\qquad Gilad Shavit\thanks{Ben-Gurion University of the Negev. \href{mailto:shavitgi@post.bgu.ac.il}{\texttt{shavitgi@post.bgu.ac.il}}. Supported by Israel Science Foundation grant \#3450/24.
}
}
\begin{document}
\pagenumbering{gobble}

\maketitle

\begin{abstract}

Thresholds for local properties tell us which guarantees to expect from random linear codes (RLCs) and help us carry those guarantees over to structured codes. A monotone property is \emph{$b$-local} if a collection of at most $b$ codewords can witness its satisfaction or failure, as in list decoding and list recovery. The threshold of $\mathcal P$ is the code dimension at which the probability that an RLC in $\F_q^n$ satisfies $\mathcal P$ crosses $1/2$ in a sharp transition. We show that the threshold of a $b$-local property $\mathcal P$ is within $O(b)$ dimensions of the \emph{expectation threshold}. This is the dimension predicted by the best union bound over \emph{covers} of $\mathcal P$: families of subspaces such that the span of every witness contains at least one member.

This gives a general derandomization statement: Any decreasing local property satisfied by RLCs at a given rate holds at a similar rate for pseudorandom ensembles that fool the cover union bound. In addition to previously studied ensembles, these include codes whose parity-check matrix is sampled from a small-bias distribution. Because our threshold characterization applies to all local properties, this derandomization also reaches properties beyond the scope of earlier reductions, including maximum loads in linear hashing for arbitrary input sets, rank-metric list decoding, and list decoding from insertion-deletion errors and burst errors.

Our theorem unifies and strengthens prior works on row-symmetric and coordinate-wise linear characterizations, which located the threshold to within $O(q^{2b}\log_q n)$ and $O(b^2+nb/\log_2 q)$ dimensions, respectively. We obtain $O(b)$ error for \emph{every} $b$-local property, independently of alphabet size. Under suitable pseudorandomness bounds, the rate loss vanishes for every $b=o(n)$, compared to $b=O(\log n)$ for earlier frameworks.

\end{abstract}

\clearpage
\tableofcontents
\clearpage
\pagenumbering{arabic}

\section{Introduction}

The theory of \deffont{local properties} allows us to analyze random codes, transfer guarantees between code ensembles, and construct codes with additional algebraic or combinatorial structure. Its recent applications include breakthroughs in explicit constructions of classical and quantum codes~\cite{BCDZ26,JS26,JMS26} and the analysis of proximity gaps~\cite{GG26,GGSW26}. A common principle drives many of these applications: guarantees are transferred from random linear codes to structured code ensembles. Hence, progress in understanding random linear codes can advance several families of code constructions at once.

Two canonical and extensively studied examples are \deffont{list-decodability} and \deffont{list-recoverability}. A code $C\subseteq\F_q^n$ is $\LD \rho L$ if every word in $\F_q^n$ lies within Hamming distance $\rho n$ of at most $L$ codewords in $C$. To certify failure, we need only exhibit $L+1$ distinct codewords in a common Hamming ball of radius $\rho n$. We therefore call $(\rho,L)$-list-decodability a \deffont{decreasing $(L+1)$-local property}. In list recovery, the receiver specifies a small list of possible symbols at each coordinate, and we require at most $L$ codewords to agree with these lists on sufficiently many coordinates. Here too, $L+1$ codewords certify failure.

More generally, a code property is \deffont{increasing} if it is upward-closed under containment. We call an increasing property $\cP$ \deffont{$b$-local} if every code satisfying $\cP$ contains a set of at most $b$ codewords whose presence certifies membership in $\cP$. A \deffont{decreasing} property is downward-closed under containment, and we call it $b$-local if its complement is $b$-local. Thus, a small collection of codewords certifies satisfaction of an increasing local property or failure of a decreasing one.

We now focus on linear codes. Fix a positive integer $n$ and a prime power $q$. For each integer $0\le k\le n$, let $C_k$ denote a \deffont{random linear code (RLC)} of dimension $k$, sampled uniformly from among the $k$-dimensional subspaces of $\F_q^n$. Its rate is $R=k/n$. All vector spaces and dimensions are over $\F_q$.
Rossman~\cite{Rossman20} proved that every nontrivial monotone code property has a sharp threshold in this model. Namely, for a nontrivial increasing property $\cP$, define its \deffont{threshold dimension} by
$$\thresholddim\inparen{\cP}:=\min\inset{k\in\inset{0,\ldots,n}:\PR{C_k\in\cP}\ge\tfrac12}\eperiod$$
Rossman's theorem implies that
\begin{align*}
\PR{C_{\thresholddim\inparen{\cP}-t}\in\cP}
&\le q^{-\Omega(t)}\ecomma\\
\PR{C_{\thresholddim\inparen{\cP}+t}\in\cP}
&\ge 1-q^{-\Omega(t)}\ecomma
\end{align*}
for every integer $t\ge2$ for which the corresponding code dimension lies between $0$ and $n$. Taking complements gives the analogous statement for decreasing properties. Rossman's theorem controls the width of the transition. To determine which code dimensions satisfy a property, we must also locate its threshold.

Mosheiff, Resch, Ron-Zewi, Silas, and Wootters~\cite{MRRSW20} characterized RLC thresholds for \deffont{row-symmetric local properties}, namely, local properties invariant under coordinate permutations, over \deffont{small alphabets}.\footnote{\cite{MRRSW20} includes coordinate-permutation invariance in its definition of ``local properties.'' We follow the later terminology, established in \cite{GMRSW21}, and call these properties \deffont{row-symmetric}.} Levi, Mosheiff, and Shagrithaya~\cite{LMS25} developed a corresponding threshold characterization for \deffont{local coordinate-wise linear (LCL)} properties in the \deffont{large-alphabet} regime. These properties are defined through witnesses consisting of tuples of codewords that satisfy prescribed linear constraints separately at each coordinate.

In~\cite{MRRSW20}, the authors also introduced a general approach to \deffont{code reductions}. They used their threshold characterization to transfer RLC guarantees to random low-density parity-check (LDPC) codes for every row-symmetric local property within its parameter range, establishing in particular that random LDPC codes achieve list-decoding capacity. Subsequent reductions transfer guarantees for the same broad class of properties to randomly punctured low-bias codes~\cite{GM22} and algebraic constructions that use substantially less randomness~\cite{MRSY25}. The reduction in~\cite{LMS25} transfers guarantees to random Reed--Solomon codes for the full class of local coordinate-wise linear properties within its scope, yielding strong list-recovery guarantees for these codes.

The coordinate-wise linear framework has also enabled explicit constructions matching RLC guarantees for broad classes of local properties. The work~\cite{BCDZ26} establishes such guarantees for subspace-design codes, including folded Reed--Solomon and univariate multiplicity codes, while~\cite{JS26} gives explicit constructions over enlarged constant alphabets. A quantum extension yields explicit quantum LDPC codes with optimal list sizes for list decoding and list recovery~\cite{JMS26}.

\subsection{Our Contribution}

We characterize the threshold of every $b$-local property to within $O(b)$ dimensions, independently of the alphabet size. We express the threshold, up to this error, in terms of the best union bound over \deffont{covers} of witnesses. A cover is a family of subspaces such that the span of every witness contains at least one member. Any linear code containing a witness must therefore contain a member of the cover, so we can bound the probability of containing a witness by taking a union bound over the cover. A cover can replace many witnesses by a subspace contained in all their spans, counting that subspace only once. 

We also exhibit a binary row-symmetric $b$-local property for which the threshold predicted by the best cover union bound lies $\log_2 b+O(1)$ dimensions below the probability threshold, showing that some dependence on locality is necessary (see \cref{sec:discussion,app:full-support-threshold-gap}).

Our cover characterization applies to all local properties, including those defined by arbitrary constraints within and across coordinates. We require neither the coordinate-permutation invariance of~\cite{MRRSW20} nor the coordinate-wise linear constraints of~\cite{LMS25}. This freedom brings properties such as \deffont{maximum loads in linear hashing} for arbitrary input sets, \deffont{rank-metric list decoding}, \deffont{list decoding from burst errors and erasures}, and \deffont{insertion--deletion list decoding} under a single threshold characterization.

The quantitative improvement is substantial even for properties covered by the earlier results. Our uniform $O(b)$ dimension error replaces the $O(q^{2b}\log_q n)$ error in~\cite{MRRSW20} and the $O(b^2+nb/\log_2 q)$ error in typical instances of \cite{LMS25}, such as list decoding (see discussion in \cref{sec:intro-previous-thresholds}). We thus obtain vanishing rate error for every locality $b=o(n)$, over every alphabet.

We also recover both earlier threshold formulas and sharpen their quantitative bounds (see \cref{cor:intro-row-threshold,cor:intro-lcl-threshold}). These explicit formulas use restricted constructions of covers, whereas our $O(b)$ characterization considers all covers. To recover the formulas, we therefore account for additional losses from restricting the covers and from estimating and combining their costs. We bound these losses through a common argument, unifying results previously established by separate methods. 

Our characterization also lets us analyze a code ensemble once and transfer RLC guarantees for an entire range of local properties. For a given locality $b$, we need only bound the probability that a sampled code contains each fixed subspace of dimension at most $b$. We call an ensemble \deffont{pseudorandom} when these probabilities satisfy the corresponding RLC upper bounds. This single condition lets us transfer guarantees for every decreasing $b$-local property. Building on the containment bounds underlying earlier reductions, we apply this principle to ensembles including \deffont{Gallager's random LDPC codes}~\cite{MRRSW20}, \deffont{randomly punctured low-bias codes}~\cite{GM22}, and \deffont{randomness-efficient algebraic constructions}~\cite{MRSY25}. We also establish the required containment bounds for \deffont{kernels of small-bias matrices}.

Beyond extending reductions to more properties, we improve their guarantees quantitatively. Suppose an RLC of dimension $k$ satisfies a decreasing $b$-local property with probability at least $1/2$. Under sufficiently strong pseudorandomness bounds, our reduction gives the same success guarantee for codes from the target ensemble at dimension $k-O(b)$, independently of $q$. %
Thus, we can replace RLCs by sufficiently pseudorandom codes while reducing their rate by only $O(b/n)$, which vanishes whenever $b=o(n)$.

Binary list decoding illustrates this improvement. If a binary RLC of dimension $k$ is $(\rho,L)$-list-decodable with probability at least $1/2$, then random punctures of a sufficiently low-bias mother code of dimension $k-O(L)$ satisfy the same list-decoding guarantee (see \cref{cor:intro-binary-list-decoding}). We preserve both the decoding radius $\rho$ and the list bound $L$, paying only for the reduction in dimension. This lets us transfer guarantees for polynomial list sizes $L=n^\alpha$, for every fixed $0<\alpha<1$, with vanishing rate loss, whereas the earlier bounds restricted this threshold comparison to logarithmic list sizes.

As another example, we use small-bias matrices to construct linear hash functions for arbitrary input sets with substantially less randomness. Let $n\ge m\ge 1$ be integers. Let $q$ be a fixed prime power, and $\delta\in (0,1)$ a fixed constant. For any input set $S\subseteq\F_q^n$ of size $q^{m(1-\delta)}$, Kumar and Mon showed that with good probability, when $S\subset \F_q^n$ is hashed into $\F_q^m$ by a random $\F_q$-linear map, the max load is $O_q(1/\delta)$ with good probability, matching the guarantees of fully random hashing \cite{KM26}. We show that small-bias matrices provide the same guarantee, but using only $O(\log n + m/\delta)$ bits of randomness, compared to $O(mn)$ for a uniformly random matrix. In the canonical setting of $n = O(m)$, this gives optimal seed length of $O_{q,\delta}(m)$ (see \cref{cor:intro-hashing}).

Neither the properties nor the ensembles above exhaust the framework. The recent use of~\cite{LMS25} to establish proximity gaps~\cite{GG26} illustrates how a general reduction can find applications beyond its original motivation. Each new RLC guarantee for a decreasing local property can benefit every applicable pseudorandom ensemble, while each new pseudorandomness bound can transfer the accumulated guarantees to another family of codes.

\subsection{Our Threshold Characterization: Covers and Expectation Thresholds}\label{sec:intro-theorem}

We locate the threshold for a given local property by optimizing a union bound over covers of \deffont{witnesses}. Let $\cA$ be a nonempty family of nonzero subspaces of $\F_q^n$, and write $\cP_\cA$ for the property that a code contains some $A\in\cA$.\footnote{The empty witness family defines a property that never holds, while a witness family containing the zero subspace defines a property that always holds. We set these degenerate cases aside throughout for brevity.} Every nontrivial increasing property of linear codes admits such a representation: we may take $\cA$ to be its inclusion-minimal members. If the property is $b$-local, these witnesses have dimension at most $b$.

A direct union bound over all witnesses may severely overestimate the probability of satisfying $\cP_\cA$ if these witness subspaces are heavily overlapping. We can reduce this overestimate by selecting a \deffont{cover}: a family $\cA'$ of subspaces such that every $A\in\cA$ contains some $A'\in\cA'$. Any code containing a witness must then contain a member of the cover. Since a fixed $d$-dimensional subspace lies in $C_k$ with probability at most $q^{-(n-k)d}$, the union bound gives
\begin{align*}
\PR{C_k\in\cP_\cA}
&\le \sum_{A'\in\cA'}\PR{A'\subseteq C_k}\\
&\le \sum_{A'\in\cA'}q^{-(n-k)\dim A'}
=: \covercost_k\inparen{\cA'}\eperiod \numberthis \label{eq:UnionBound}
\end{align*}
We call $\covercost_k\inparen{\cA'}$ the \deffont{cost} of the cover at dimension $k$. A cover of cost at most $\frac12$ certifies that the random code contains no witness with probability at least $\frac12$.

Let $\covers(\cA)$ denote the family of all covers of $\cA$. We define the \deffont{expectation threshold} by
\begin{equation}\label{eq:ExpDim}
\expectationdim\inparen{\cP_\cA}
:=
\max_{\cA'\in\covers(\cA)}
\min\inset{k\in\inset{0,\ldots,n}:
\covercost_k\inparen{\cA'}\ge\tfrac12}\eperiod
\end{equation}
For each cover, the inner minimum gives the first dimension at which its cost reaches $1/2$. We maximize this dimension over all covers to obtain the strongest union-bound guarantee. Equivalently, the expectation threshold is the smallest dimension at which every cover has cost at least $1/2$. Our main theorem shows that optimizing this union bound locates the probability threshold to within $O(b)$ dimensions.

\begin{theorem}[Threshold characterization]\label{thm:intro-threshold}
Let $q$ be a prime power, let $n,b$ be positive integers, and let $\cP$ be a nontrivial increasing $b$-local property of linear codes in $\F_q^n$. Then
$$
\expectationdim\inparen{\cP}
\le \thresholddim\inparen{\cP}
\le \expectationdim\inparen{\cP}+2b+2\eperiod
$$
\end{theorem}
The first inequality follows immediately from the union bound in \cref{eq:UnionBound}. The core of the theorem is the second inequality, which asserts that some cover certifies nonoccurrence almost up to the probability threshold. We prove the theorem in \cref{sec:threshold-proof}.

For a nontrivial decreasing property $\cP$, we define $\thresholddim\inparen{\cP}$ as the smallest dimension at which an RLC satisfies $\cP$ with probability at most $1/2$, and define $\expectationdim\inparen{\cP}$ using witnesses to failure. These thresholds equal those of the increasing property of failing $\cP$, so the same inequalities apply.

Our characterization follows the expectation-threshold paradigm of Kahn and Kalai~\cite{KK07}. They proposed that optimizing a first-moment bound over covers should approximately locate the threshold of every increasing family of sets. Park and Pham~\cite{PP23} proved this conjecture: when a random set includes each ground-set element independently with probability $p$, the probability threshold exceeds the expectation threshold by at most a factor logarithmic in the largest minimal witness size. Their theorem applies to arbitrary increasing set families, without requiring small witnesses.

Friedgut, Kahn, and Shikhelman~\cite{FKS15} proved a closer analogue of our result for families with bounded witness size---a constraint analogous to our notion of locality. They relate the probability of an increasing set property to the minimum cost of a \deffont{fractional cover}, namely, an assignment of nonnegative weights to sets such that every witness contains sets of total weight at least one. We adapt their linear programming duality and weighted second-moment argument to subspaces.

\paragraph{Proof idea.}
We sketch our proof of the inequality $$\thresholddim\inparen{\cP}
\le \expectationdim\inparen{\cP}+2b+2\eperiod$$
We start at a dimension $k$ where an RLC contains a witness with probability less than $1/2$. Our goal is to find a cover whose cost falls below $1/2$ after decreasing the dimension by only $O(b)$. To do so, we first find a fractional cover of bounded cost, then round it to an ordinary cover.\footnote{The proof in \cref{sec:threshold-proof} handles the difference between $\PR{A\le C_k}$ and $q^{-(n-k)\dim A}$ using a slightly different normalization and a direct second-moment bound. We omit these technical adjustments from the overview for simplicity.}

Let $\cV$ consist of all nonzero subspaces contained in at least one witness, and put $s:=b+1$. We extend the cover cost function to fractional covers by multiplying each subspace's contribution by its fractional weight. We then study the linear program for a minimum-cost fractional cover of $\cA$ at dimension $k-s$:
\begin{align*}
\lambda=\min_{x_U\ge0}\quad
&\sum_{U\in\cV}x_Uq^{-(n-k+s)\dim U}\ecomma\\
\text{subject to}\quad
&\sum_{\substack{U\in\cV\\U\le A}}x_U\ge1
\qquad\text{for every }A\in\cA\eperiod
\end{align*}
An ordinary cover corresponds to a $\inset{0,1}$ variable assignment in this program.

The dual program has a nonnegative variable $y_A$ for each witness $A\in\cA$:
\begin{align*}
\lambda=\max_{y_A\ge0}\quad
&\sum_{A\in\cA}y_A\ecomma\\
\text{subject to}\quad
&\sum_{\substack{A\in\cA\\U\le A}}y_A
\le q^{-(n-k+s)\dim U}
\qquad\text{for every }U\in\cV\eperiod
\end{align*}
Each constraint limits the total weight of witnesses containing a fixed subspace. 

An optimal dual solution defines a nonnegative linear combination of witness indicators:
$$
X:=
\sum_{A\in\cA}
\frac{y_A}{\PR{A\le C_k}}
\mathbf{1}_{\{A\le C_k\}}\eperiod
$$
The normalization gives $\E{X}=\sum_Ay_A=\lambda$, while $X>0$ certifies that the code contains a witness.

We next bound the variance of $X$. Our goal is to show that, if the dual optimum $\lambda=\E{X}$ were large, then $X$ would concentrate around its positive mean.

We analyze the covariance between pairs of witness indicators. Witnesses with trivial intersection contribute no positive covariance. For the remaining pairs, we group the contributions according to the intersection $U$. The dual constraint for $U$ bounds the total weight of witnesses containing that subspace.

Here we use locality for the first time. A witness of dimension at most $b$ contains at most $q^{ba}$ subspaces of dimension $a$. The decrease by $s=b+1$ dimensions contributes a factor $q^{-sa}$ when we apply the dual constraint for each such subspace. Thus, summing the contributions of the possible intersections gives a bound proportional to
$$
\sum_{a=1}^b q^{ba}q^{-(b+1)a}
=\sum_{a=1}^b q^{-a}<1\eperiod
$$
This bound is uniform over $q$, and yields
$$
\operatorname{Var}(X)=O\inparen{\E{X}}\eperiod
$$
If $\lambda$ were sufficiently large, the variance would be much smaller than $\E{X}^2$. The variable $X$ would then concentrate around its positive mean, forcing the code to contain a witness with probability greater than $1/2$. This contradicts our choice of $k$. Thus, the dual optimum is bounded by an absolute constant, and linear programming duality supplies a fractional cover of bounded cost.

We use locality a second time to round this fractional cover. We retain subspaces whose weights exceed a suitable dimension-dependent cutoff. Because each witness has dimension at most $b$, we can choose these cutoffs so that every witness contains a retained subspace. Decreasing the code dimension by another $b+1$ compensates for replacing the retained fractional weights by one. The resulting ordinary cover has cost less than $1/2$. Together, the two steps give the dimension decrease $2b+2$.

The second-moment argument in~\cite{MRRSW20} essentially analyzes an unweighted version of $X$. For witness families with the structure described in \cref{sec:earlier-threshold-proofs}, this unweighted variable already has sufficiently small variance to yield the needed concentration bound. The freedom to choose weights through the dual program lets us adapt $X$ to the overlaps among witnesses and obtain the variance bound for arbitrary $b$-local witness families, without these structural assumptions. The freedom to choose weights through the dual program lets us adapt $X$ to the overlaps among witnesses and obtain the variance bound for arbitrary $b$-local witness families, without these structural assumptions.

\subsection{Local Properties Beyond Coordinate Symmetry}\label{sec:intro-properties}

We use the following examples to illustrate both the scope of our threshold characterization and concrete applications of our reductions. The properties have failure witnesses that depend on arbitrary input sets, coordinate locations, or relations between coordinates. Our theorem characterizes their RLC thresholds through covers of these witnesses. Alongside the witness descriptions, we recall known RLC guarantees that we can now transfer to ensembles with suitable pseudorandomness parameters (see \cref{sec:intro-reductions}). All the properties below are decreasing, and $L\ge1$ denotes the list size bound.

\paragraph{Linear hashing and arbitrary error sets.}
In linear hashing, we use a linear map $h:\F_q^n\to\F_q^m$ to distribute a prescribed input set $S\subset \F_q^n$ among $M\coloneqq q^m$ bins. We seek to control the maximum load, namely, the largest number of inputs assigned to one bin. This quantity satisfies
$$
\max_{y\in\F_q^m}|S\cap h^{-1}(y)|
=\max_{z\in\F_q^n}|\ker h\cap(z+S)|\eperiod
$$ When $S$ is a Hamming ball of radius $\rho n$, the map $h$ has maximum load at most $L$ on $S$ if and only if $\ker h$ is $(\rho,L)$-list-decodable (see~\cite{dlmnrr,KM26}). For general $S$, maximum load at most $L$ is equivalent to \deffont{$(S,L)$-list-decodability} of $\ker h$: every translate of $S$ contains at most $L$ codewords. For a linear code, this condition guarantees list size at most $L$ against adversarially chosen additive errors from $S$.
 This property is $L$-local: we can translate an offending list to contain zero, leaving only $L$ words to specify.

For constant $q$, it was recently shown that a uniformly random surjective linear map has maximum load matching that of a fully random map with high probability~\cite[Theorem~3.2]{KM26}. The kernel of this random map is an RLC of dimension $n-m$. 

In \cref{cor:intro-hashing}, we replace the uniformly random hash matrix by one sampled from a small-bias distribution, retaining the maximum-load bound with $m+o(m)$ output symbols and substantially fewer random bits. For example, put $M=q^m$ and consider an arbitrary input set $S$ of size $M^{1-\eps}$, where $0<\eps<1$ is fixed. A fully random function maps these inputs into $M$ bins with maximum load $O(1/\eps)$ with high probability~\cite{RS98-ballsbinstightanalysis}. Our construction achieves the same asymptotic load bound with high probability with a pseudorandom linear map that only uses $O(\eps^{-1}\log M+\log n)$ random bits. A comparable seed length can be obtained by standard constructions, such as almost $O(1/\eps)$-wise independence. However, we are interested in preserving linearity of our hash functions, which is a desirable property crucially used in cryptography and algorithms~\cite{BGG94,ams96,abf23}. We thus obtain a derandomized family that matches the load guarantee of fully random hashing, while simultaneously preserving linearity.

\paragraph{Rank-metric list decoding.}
Rank-metric decoding treats codewords as matrices and allows errors of bounded rank. When $n=uv$, we identify $\F_q^n$ with $\F_q^{u\times v}$ and measure distance by the rank of the difference. We normalize the decoding radius by $\min\inset{u,v}$ when referring to a relative radius. We require every ball of the prescribed decoding radius to contain at most $L$ codewords. Translating an offending list by one of its $L+1$ members leaves $L$ differences that still certify failure, proving $L$-locality. For fixed alphabets and relative decoding radii, RLCs approach rank-metric list-decoding capacity as both matrix dimensions grow, with list size proportional to the inverse rate gap~\cite{GR18}.

\paragraph{List decoding from burst errors and erasures.}
A burst corrupts or erases symbols within an interval of at most $t$ consecutive coordinates, with erasure positions revealed to the receiver. Translating an offending list to contain zero proves $L$-locality. For pure erasures, a nonempty decoding list is an affine subspace whose direction space consists of codewords supported on the erased positions. A list larger than $L$ therefore yields $\lfloor\log_q L\rfloor+1$ linearly independent differences supported on the erased positions. These differences certify failure and give a smaller locality bound. 

These properties need not be row-symmetric, although the coordinate-wise linear framework of~\cite{LMS25} captures them by fixing the candidate burst intervals. Our characterization opens a larger range of list sizes: for burst-error list decoding, the earlier general threshold bounds give vanishing rate error when $L=o(\sqrt n)$, whereas our cover characterization gives error $O(L/n)$ over every alphabet and therefore handles every $L=o(n)$.

An elementary covering argument gives strong RLC guarantees for all these models. For every fixed $\varepsilon>0$ and $1\le t\le(1-\varepsilon)n$, RLCs of rate at most $1-t/n-\varepsilon$ have list size $O(1/\varepsilon)$ with high probability as $n\to\infty$, over every alphabet.\footnote{Group burst starting positions into blocks of $\lceil\varepsilon n/2\rceil$ consecutive positions. Each group fits inside an interval of length less than $t+\varepsilon n/2$. With high probability, the code contains no nonzero word supported on any such interval, so each group contributes at most one candidate. Filling erased positions arbitrarily reduces mixed bursts to the error-only case.}

\paragraph{Insertion--deletion list decoding.}
An insertion--deletion channel inserts or removes symbols, shifting the alignment between transmitted and received words. We require every received word to arise from at most $L$ codewords within a prescribed edit budget. The $L+1$ offending codewords certify failure, proving $(L+1)$-locality.

RLCs support unique decoding near the half-Singleton bound over large alphabets~\cite{CGHL21} and binary list decoding at rates approaching one as the error fraction vanishes~\cite[Lemma~11]{LGF25}. The binary list-decoding result thus provides an input to our reductions for a channel that changes coordinate alignments.

\subsection{Reductions to Pseudorandom Code Ensembles}\label{sec:intro-reductions}

We now use our threshold characterization to transfer RLC guarantees to structured code ensembles. We measure the required pseudorandomness through subspace containment: for every fixed subspace $U$ of dimension at most $b$, the target code must contain $U$ with probability at most $q^{-(n-k)\dim U}$, matching the upper bound for an RLC of dimension $k$.

The idea behind the reductions is to use the same cover to bound failure for both RLCs and the target ensemble. For every decreasing $b$-local property, \cref{thm:intro-threshold} guarantees a cover of its failure witnesses with cost below $1/2$ when $k$ lies only $O(b)$ dimensions below the RLC threshold. Any code that fails the property contains a member of this cover. If the target ensemble satisfies the containment bounds, the union bound over the same cover therefore shows that the target code satisfies the property with probability at least $1/2$. 

Concretely, for fixed $q$ and $n$, a \deffont{code ensemble} is a probability distribution on linear subspaces of $\F_q^n$. Some ensembles obey the required containment bounds only after we exclude a rare event, such as unusually small minimum distance. The following definition includes this possibility.

\begin{definition}[Pseudorandom code ensembles]\label{def:intro-pseudorandom}
Let $b\ge1$ be an integer and let $R\in[0,1]$. We call an ensemble \deffont{\pseudorandom{b}{R}} if a code $C$ sampled from it satisfies
$$
\PR{U\le C}\le q^{-n(1-R)\dim U}
\qquad\text{for every }U\le\F_q^n\text{ with }1\le\dim U\le b\eperiod
$$
For $\delta\in[0,1]$, we call the ensemble \deffont{\almostpseudorandom{b}{R}{\delta}} if there is a single event $G$ in the sampling experiment, of probability at least $1-\delta$, such that
$$
\PR{U\le C\text{ and }G}\le q^{-n(1-R)\dim U}
\qquad\text{for every }U\le\F_q^n\text{ with }1\le\dim U\le b\eperiod
$$
Taking $\delta=0$ recovers ordinary pseudorandomness.

We call $R$ the \deffont{comparison rate}: when $nR$ is an integer, an RLC of dimension $nR$ satisfies the same upper containment bound. We do not require the sampled codes to have dimension $nR$.
\end{definition}

\begin{corollary}[Code reductions]\label{cor:intro-transfer}
Let $\cP$ be a nontrivial decreasing $b$-local property. Suppose that $C$ comes from an \almostpseudorandom{b}{R}{\delta} ensemble. Then, for every real $t\ge0$,
$$
nR\le \thresholddim\inparen{\cP}-2b-4-t
\quad\Longrightarrow\quad
\PR{C\notin\cP}\le\delta+\frac12q^{-t}\eperiod
$$
\end{corollary}

\begin{remark}[Quantitative improvement over earlier reductions]\label{rem:intro-reduction-parameters}
For row-symmetric properties, reductions based on~\cite{MRRSW20} incur a dimension loss of $O(q^{2b}\log_q n+t)$ to obtain failure probability at most $q^{-t}$ under \pseudorandom{b}{R} containment bounds~\cite[Proposition~2.16]{MRSY25}. Under the same containment bounds, \cref{cor:intro-transfer} requires only $2b+4+t$ dimensions to obtain failure probability at most $\tfrac12q^{-t}$, and applies to every decreasing $b$-local property.
\end{remark}

We prove \cref{cor:intro-transfer} in \cref{app:transfer-proof}.

The reduction principle provides a general route to derandomization: an ensemble that uses fewer random bits can inherit RLC guarantees for every decreasing local property within its pseudorandomness range. For example, sampling the linearized-polynomial ensemble of~\cite{MRSY25} uses $O(bn\log q)$ random bits. For fixed alphabet size and locality, this is $O(n)$, compared with $\Theta(n^2)$ for an RLC at a fixed rate in $(0,1)$. Over a fixed alphabet, the cost remains subquadratic whenever $b=o(n)$.

Another simple construction samples all entries of a parity-check matrix jointly from a small-bias distribution. We call a random matrix $A$ \deffont{$\varepsilon$-biased} if
$$
\left|\E{\chi(\varphi(A))}\right|\le\varepsilon
$$
for every nonzero $\F_q$-linear form $\varphi$ in its entries and every nontrivial additive character $\chi$ of $\F_q$. Over $\F_2$, this means that every nontrivial linear functional of the entries has bias at most $\varepsilon$.

For $A\in\F_q^{n\times r}$ with $1\le r\le n$, we take $C:=\ker A^{\mathsf T}$. If $\varepsilon\le q^{-rb}$, this ensemble is \pseudorandom{b}{1-(r-1)/n}. Every sampled code has dimension at least $n-r$, so the comparison dimension is only one larger than this guaranteed dimension. Small-bias constructions~\cite{NN90,AGHP90} sample such matrices using $O(\log_2(nr)+rb\log_2 q)$ random bits. We prove the containment bound and derive the randomness bound in \cref{app:small-bias-matrices}.

Random puncturing offers a complementary benefit: we can retain structure from a chosen mother code while obtaining RLC guarantees. Puncturing selects columns of the mother code's generator matrix, preserving their algebraic descriptions. A sufficiently low-bias mother code\footnote{A linear code $D\le\F_q^N$ has bias at most $\beta$ if $|N^{-1}\sum_{i=1}^N\chi(c_i)|\le\beta$ for every nonzero $c\in D$ and every nontrivial additive character $\chi$ of $\F_q$.} supplies the pseudorandomness needed for the reduction. Expander-walk sampling~\cite{PP23b} further reduces the randomness needed to choose the coordinates. We give the parameters in \cref{app:low-bias-punctures}.

We collect the pseudorandomness guarantees for low-bias punctures, linearized-polynomial codes, kernels of small-bias matrices, and Gallager's random LDPC codes in \cref{cor:intro-ensembles}, stated in \cref{app:ensemble-guarantees}. For low-bias punctures, linearized-polynomial codes and Gallager's ensemble, we build on the containment bounds of~\cite{GM22,MRSY25,MRRSW20}.

The next two corollaries illustrate the improved dimension loss and the wider range of properties covered by our reductions. The first uses low-bias punctures to preserve a prescribed binary Hamming list bound $L$ at a dimension only $2L+6$ below its RLC threshold. The second uses a small-bias matrix to construct hash maps for arbitrary input sets. Both applications follow from \cref{cor:intro-transfer}.

\begin{corollary}[Binary list decoding from low-bias punctures]\label{cor:intro-binary-list-decoding}
Let $\cP$ denote $(\rho,L)$-list-decodability for binary linear codes of length $n$, where $L\ge1$. Let $D\le\F_2^N$ have dimension $k$ and bias at most $2^{-L}/(2n)$. For every $t\ge0$, if
$$
k\le\thresholddim\inparen{\cP}-2L-6-t\ecomma
$$
then a random puncturing of $D$ to $n$ coordinates, sampled independently and uniformly with replacement, has dimension $k$ and satisfies $\cP$ with probability at least $1-\tfrac12 2^{-t}$.
\end{corollary}

We next apply the small-bias-matrix ensemble to hashing arbitrary input sets. As explained in \cref{sec:intro-properties}, the map $h(x)=A^{\mathsf T}x$ has maximum load at most $L$ on $S$ exactly when $\ker A^{\mathsf T}$ is $(S,L)$-list-decodable. Since this is a decreasing $L$-local property, we can use the kernel ensemble's pseudorandomness and \cref{cor:intro-transfer} to transfer the RLC maximum-load guarantee to small-bias matrices.

\begin{corollary}[Hashing from small-bias matrices]\label{cor:intro-hashing}
Fix $q$ and choose a sufficiently large constant $c_q$. Let $m\ge2$ be an integer, let $S\subseteq\F_q^n$ have size $1\le N\le q^m$, and set
$$
L:=\left\lceil
\frac{c_qm}{m-\log_qN+\log_qm}
\right\rceil\eperiod
$$
Let $t\ge0$ be an integer and suppose that
$$
r:=m+2L+5+t\le n\eperiod
$$
Sample an $\varepsilon$-biased matrix $A\in\F_q^{n\times r}$ with $\varepsilon\le q^{-rL}$, and define
$$
h:\F_q^n\longrightarrow\F_q^r,
\qquad h(x):=A^{\mathsf T}x\eperiod
$$
For all sufficiently large $m$, with probability at least $1-\tfrac12q^{-t}$, the map $h$ has maximum load at most $L$ on $S$. Equivalently, $\ker h$ is $(S,L)$-list-decodable.

An explicit choice of the distribution of $A$ uses $O(\log_2(nr)+rL\log_2q)$ random bits. Choosing also $t\to\infty$ makes the failure probability vanish.
\end{corollary}

 If $t=0$, $n = \Theta(m)$, and $q = O(1)$, then $r=m+2L+5$ and the pseudorandom hash maps into $q^{m+2L+5}$ bins with seed length $O(mL)$. One can compute (say, with \cite{RS98-ballsbinstightanalysis}) that for $N \le q^{m-\sqrt{m}}$, $L$ matches the expected max load when $N$ balls are tossed into $q^{m+2L+5}$ bins up to constant factors. Therefore, we obtain a derandomization of linear hashing when tossing $N$ balls into $M\ge N e^{\Omega(\log^{1/2} N)}$ bins, and obtain optimal seed length of $O(m)$ when $M\ge N^{1+\delta}$ for some constant $\delta > 0$.

yWe prove \cref{cor:intro-binary-list-decoding} in \cref{app:binary-list-decoding}, where we also give its expander-walk variant, and prove \cref{cor:intro-hashing} in \cref{app:hashing}.

\subsection{A Unified View of Earlier Threshold Characterizations}\label{sec:intro-previous-thresholds}

We refer to the frameworks of~\cite{MRRSW20} and~\cite{LMS25} as the \deffont{row-symmetric setting} and the \deffont{local coordinate-wise linear (LCL) setting}, respectively. Each work characterizes RLC thresholds for local properties in its setting. Our main theorem, \cref{thm:intro-threshold}, characterizes the threshold through an optimization over all covers of the witnesses. We can view the earlier row-symmetric and LCL characterizations as using the same optimization over only a restricted collection of covers. Here we explain which covers they use and why those covers suffice to characterize the threshold in the respective settings. In \cref{sec:earlier-threshold-proofs}, we state both earlier formulas with improved quantitative bounds, give a tighter version of the row-symmetric formula, and derive these results from our main theorem.

We work with increasing properties; for a decreasing property, we apply the discussion to its failure witnesses. To explain the earlier characterizations, we first describe witnesses using matrices.
Suppose that the presence of codewords $x_1,\ldots,x_d$ certifies that a code satisfies the property. Place these codewords in the columns of a matrix $M\in\F_q^{n\times d}$. Since the code is linear, it contains these columns if and only if it contains their span, $\operatorname{col}(M)$. Thus, $M$ represents a witness subspace. We call $M$ a \deffont{witness matrix}. Its rows record the symbols of the witness codewords at each coordinate. Different witness matrices may represent the same subspace.

Both earlier works organize witness matrices into families with common structure. In the row-symmetric setting, each family consists of matrices with the same multiset of rows. In the LCL setting, a \deffont{profile} specifies the allowed linear constraints at every coordinate. The authors group witness matrices by their profile and by the linear relations among their columns.

Within each family, the matrices have exactly the same linear relations among their columns. We therefore retain the same subset of columns in every matrix to form a basis of its witness subspace. Let $\cS_j$, indexed by $j\in J$, denote the resulting nonempty families of basis matrices. For each $j$, let $\cP_j$ be the property that a code contains the columns of some matrix in $\cS_j$. The original property then satisfies
$$
\cP=\bigcup_{j\in J}\cP_j\eperiod
$$
Thus, a code satisfies $\cP$ exactly when it contains a witness from at least one of the families. We first analyze each family separately, then account for the cost of combining their covers.

\paragraph{Interpreting the earlier characterizations through covers.}
Fix one family $\cS_j$ and write $\cS:=\cS_j\subseteq\F_q^{n\times d}$, where every matrix has rank $d\le b$. Choose a rank-$a$ coefficient matrix $B\in\F_q^{d\times a}$, where $1\le a\le d$, and apply it to every $M\in\cS$. The spaces
$$
\cA'_B:=\inset{\operatorname{col}(MB):M\in\cS}
$$
form a cover of the witness subspaces represented by $\cS$, since each witness $\operatorname{col}(M)$ contains the corresponding cover member $\operatorname{col}(MB)$. Every cover member has dimension $a$. Writing $\cS B:=\inset{MB:M\in\cS}$, we can therefore bound its cost by
$$
\covercost_k(\cA'_B)\le|\cS B|q^{-(n-k)a}\eperiod
$$
Several projected matrices may span the same subspace, so counting matrices can overestimate the cover cost.

We define the \deffont{projection-expectation threshold} by restricting the optimization in \cref{eq:ExpDim} to the \deffont{common-projection covers} $\cA'_B$:
$$
\projectionexpectationdim\inparen{\cS}
:=
\max_{\substack{1\le a\le d\\
B\in\F_q^{d\times a},\ \rank B=a}}
\min\inset{k\in\inset{0,\ldots,n}:
\covercost_k\inparen{\cA'_B}\ge\tfrac12}\eperiod
$$
The cover cost counts each distinct subspace in $\cA'_B$ once. We optimize over the coefficient matrix $B$ and the selected dimension $a$, but must use the same choice for every witness matrix in $\cS$. 

The earlier works count projected matrices using entropy in the row-symmetric setting and dimensions in the LCL setting. In our terminology, they use these counts to estimate the projection-expectation threshold. In the unrestricted cover optimization, however, we may choose both the coefficient matrix and the selected dimension separately for each witness.

Restricting the collection of covers can only decrease the maximum in \cref{eq:ExpDim}. Together with \cref{thm:intro-threshold}, this gives
$$
\projectionexpectationdim\inparen{\cS}
\le\expectationdim\inparen{\cP_j}
\le\thresholddim\inparen{\cP_j}
\le\expectationdim\inparen{\cP_j}+2b+2\eperiod
$$
Our main theorem places the unrestricted expectation threshold within $O(b)$ dimensions of the probability threshold. It does not, however, bound how far the projection-expectation threshold can lie below them.

For the structured matrix families arising in the row-symmetric and LCL settings, we can also show that the projection-expectation threshold is close to the probability threshold. We now explain which features of these families make common projections sufficient.

\paragraph{Why the restricted optimization suffices in the earlier settings.}
We justify the use of a common projection by studying the distribution of $MB$ when we sample $M$ uniformly from a matrix family $\cS$. We call $\cS$ \deffont{projection-uniform} if, for every fixed projection $B$, the matrix $MB$ is uniform over $\cS B$. For a projection-uniform family $\cS_j$ in the decomposition, we prove in \cref{sec:projection-principle} that
$$
\projectionexpectationdim\inparen{\cS_j}
\le\expectationdim\inparen{\cP_j}
\le\projectionexpectationdim\inparen{\cS_j}+O(b)\eperiod
$$
Within this family, allowing a separate projection for each witness therefore improves the expectation threshold by at most $O(b)$ dimensions.

In the row-symmetric setting, we use coordinate permutations to establish projection-uniformity for each $\cS_j$. In the LCL setting, we derive a similar comparison from the equal-sized fibers of linear maps. Combining these comparisons with \cref{thm:intro-threshold} gives
$$
\projectionexpectationdim\inparen{\cS_j}
\le\thresholddim\inparen{\cP_j}
\le\projectionexpectationdim\inparen{\cS_j}+O(b)\eperiod
$$
We can therefore locate the threshold of each $\cP_j$ by optimizing over common projections of its matrix family $\cS_j$.

To pass from the individual properties $\cP_j$ to the original property $\cP$, we unite the covers chosen for the matrix families $\cS_j$. We bound the cost of this union by the sum of the individual cover costs. Decreasing the code dimension by $\lceil\log_q|J|\rceil$ compensates for this sum (see \cref{lem:witness-decomposition}). The usefulness of the earlier formulas therefore depends on how many families their decompositions require. In the row-symmetric setting, the number of row distributions grows rapidly with $q$, favoring small alphabets. In typical LCL applications, the profile-count bound grows exponentially with $n$ but does not depend on $q$. Increasing $q$ then reduces the base-$q$ logarithm of that bound, favoring large alphabets.

\paragraph{The advantage of arbitrary covers.}
The earlier characterizations use structure to justify choosing one common projection for each matrix family. Without that structure, a common projection can be much less effective than an arbitrary cover. We exhibit a binary $2$-local property for which the projection-expectation threshold lies $\Theta(n)$ dimensions below the probability threshold, regardless of how we choose bases for its witness subspaces. On the other hand, our unrestricted cover characterization locates the probability threshold within a constant number of dimensions.

\begin{example}\label{ex:intro-common-projection}
Let $n\ge8$ be divisible by four, and fix $0\ne v\in\F_2^n$. Choose a family $\cD$ of $2^{n/2}$ binary planes that intersect pairwise only at zero and none of which contains $v$. Put
$$
\cA:=\inset{A\le\F_2^n:\dim A=2,\ v\in A}\cup\cD\eperiod
$$
For each witness in $\cA$, choose one or more ordered bases, and let $\cS\subseteq\F_2^{n\times2}$ consist of the resulting matrices. Regardless of these choices,
\begin{align*}
\projectionexpectationdim(\cS)&=\frac n2\ecomma\\
\expectationdim(\cP_\cA)&=\frac{3n}{4}\ecomma\\
\thresholddim(\cP_\cA)&=\frac{3n}{4}+O(1)\eperiod
\end{align*}

The gap between $\projectionexpectationdim(\cS)$ and $\expectationdim(\cP_\cA)$ arises because, in the unrestricted cover optimization, we may select a line from some witnesses and keep other witness planes in full. A common projection forces us either to select a line from every witness or to keep every witness plane.  We explain this mechanism here and defer the construction of $\cD$ and the verification of the two expectation thresholds to \cref{app:common-projection-example}.

In the unrestricted setting, consider the cover $\inset{\langle v\rangle}\cup\cD$. We select the shared line from the planes containing $v$ and keep each plane in $\cD$, giving total cost
$$
2^{k-n}+2^{n/2}2^{2(k-n)}\eperiod
$$
The unrestricted cover can therefore remain inexpensive at dimensions where every common-projection cover already has large cost. Once we establish the two expectation thresholds, the probability-threshold estimate follows from \cref{thm:intro-threshold}.
\end{example}

\subsection{Open Problems}\label{sec:discussion}

\paragraph{The optimal threshold gap.}
Our theorem bounds the gap between the probability and expectation threshold dimensions by $O(b)$, uniformly over the alphabet size. What is the optimal dependence on the locality?

A binary example shows that some dependence on $b$ is unavoidable. Let $n=2^b-1$, and let $\cP_b$ be the property of having full coordinate support: for every coordinate, some codeword is nonzero there. This property has locality exactly $b$, and, for all sufficiently large $b$,
$$
\thresholddim\inparen{\cP_b}=b+1,
\qquad
\expectationdim\inparen{\cP_b}=b-\log_2 b+O(1)\eperiod
$$
The threshold gap is therefore $\log_2 b+O(1)$ (see \cref{app:full-support-threshold-gap} for the proof).

Can we match this logarithmic dependence with a general upper bound? More precisely, does every nontrivial $b$-local monotone property satisfy
$$
\thresholddim\inparen{\cP}
-\expectationdim\inparen{\cP}
=O\inparen{1+\log_q b}
$$
uniformly over $q$? Such a bound would be optimal in its dependence on $b$ over $\F_2$. Since every minimal witness in $\F_q^n$ has dimension at most $n$, it would also locate the threshold of every nontrivial monotone property to within $O(1+\log_q n)$ dimensions, without any locality assumption.

\paragraph{Reduction to random Reed--Solomon codes.}
A Reed--Solomon code of dimension $k$ consists of all words $\inparen{f(\alpha_1),\ldots,f(\alpha_n)}$, where $\alpha_1,\ldots,\alpha_n\in\F_q$ are distinct evaluation points and $f\in\F_q[X]$ ranges over all polynomials of degree less than $k$. Choosing the evaluation points uniformly without replacement gives the random Reed--Solomon ensemble. The reduction in~\cite{LMS25} transfers RLC guarantees to random Reed--Solomon codes for LCL properties over sufficiently large alphabets. Can we extend this connection to natural classes of local properties whose witnesses couple different coordinates, such as insertion--deletion list decoding?

For comparison rates $R<1$, our pseudorandomness condition fails for standard Reed--Solomon codes of positive dimension because they all contain the line of constant words. The number of codes in the ensemble gives another obstruction. A \pseudorandom{b}{R} ensemble of $k$-dimensional codes, with $1\le b\le k$, must contain at least $q^{n(1-R)b}$ distinct codes: the probability of sampling any particular code is at most the containment bound for a $b$-dimensional subspace of that code. On the other hand, the $n$ evaluation points determine a Reed--Solomon code, leaving at most $q^n$ possibilities. These bounds are incompatible when $b(1-R)>1$.

The proof in~\cite{LMS25} shows how a comparison can instead exploit the structure of the property. For each profile, the authors analyze tuples of polynomials whose evaluation vectors satisfy its coordinate-wise constraints. As they reveal the evaluation points, these constraints restrict the possible polynomial tuples, and the proof tracks the dimensions of the remaining solution spaces. Can we extend this analysis to constraints involving several evaluation points, or develop another reduction principle that transfers RLC guarantees for such properties?

\paragraph{Derandomizing Linear Hashing.}
Let $h: \F_2^{2m}\to \F_2^m$ be a random linear map. It was shown by Jaber, Kumar, and Zuckerman \cite{JKZ25} that for any subset $S\subset \F_2^{2m}$ of size $2^m$, the max load when $h$ hashes $S$ is $O(m/\log m)$ with good probability. This uses $O(m^2)$ bits of randomness. By subsampling $2^{O(m)}$ of the $2^{2m^2}$ possible matrices, one can show that there exists a non-explicit family of pseudorandom linear maps achieving max load $O(m/\log m)$ that uses only $O(m)$ bits of randomness. Can we obtain an explicit such family of linear maps? Even obtaining a seed of $O(m^{1.99})$ would be interesting. Our techniques only obtain some randomness reduction when $|S|\ll q^m$. Proving that $\thresholddim\inparen{\cP}
-\expectationdim\inparen{\cP}
=O\inparen{1+\log_q b}$ would imply a $O(m^2/\log m)$-seed construction using small-bias matrices.

Even ignoring linearity, it is an interesting open question to obtain a hash family using $O(m)$ bits and achieves max load $O(m/\log m)$. Currently, the best max load we can achieve with $O(m)$ bits of randomness is only $2^{O(m)}$. A simple construction providing this is $C$-wise independent hashing, which achieves max load $2^{O(m/C)}$ with seed $O(Cm)$. If we are willing to sacrifice optimal seed length, hash functions achieving optimal max load with $O(m\log m)$ bits of randomness are known \cite{crsw}.

\section{Proof of the Threshold Characterization Theorem}\label{sec:threshold-proof}

We prove \cref{thm:intro-threshold} using the arguments outlined in \cref{sec:intro-theorem}, starting here from the expectation threshold. We first use rounding to lower-bound the minimum fractional-cover cost at an intermediate dimension. We then use the dual weights in a second-moment argument to show that an RLC $2b+2$ dimensions above the expectation threshold contains a witness with probability greater than $1/2$.

\begin{proof}[Proof of \cref{thm:intro-threshold}]
The inequality $\expectationdim(\cP)\le\thresholddim(\cP)$ follows from \cref{eq:UnionBound}. For the upper bound on $\thresholddim(\cP)$, put
$$
e:=\expectationdim(\cP),
\qquad h:=b+1,
\qquad k:=e+2h\eperiod
$$
If $k\ge n$, then the claim holds since $\thresholddim(\cP)\le n\le k$, so assume $k<n$. We need to show that $\PR{C_k\in\cP}>1/2$.

Let $\cA$ be the family of inclusion-minimal witnesses of $\cP$, and let $\cV$ consist of all nonzero subspaces contained in its members. Every member of $\cA$ has dimension at most $b$.

\paragraph{Bounding the contribution of a witness's subspaces.}
We will use locality in both the rounding and second-moment arguments through the sum
$$
\tau:=\sum_{d=1}^b\binom{b}{d}_{\!q}q^{-hd}\ecomma
$$
where $\binom{b}{d}_{\!q}$ counts the $d$-dimensional subspaces of $\F_q^b$. For every $A\in\cA$ we have $\dim A \le b$, yielding
\begin{align*}
\sum_{\{0\}<U\le A}q^{-h\dim U}
\le\tau\eperiod
\numberthis\label{eq:witness-sum}
\end{align*}

We need only the bound $\tau<3/5$. To prove it, write
$a_d:=\binom{b}{d}_{\!q}q^{-hd}$. Then $a_1<1/2$, and for $1\le d<b$,
$$
\frac{a_{d+1}}{a_d}
=q^{-(b+1)}\frac{q^{b-d}-1}{q^{d+1}-1}
<\frac1{q^{d+1}(q^{d+1}-1)}
\le\frac1{12}\eperiod
$$
Thus,
$$
\tau<\frac12\sum_{j\ge0}12^{-j}
=\frac6{11}<\frac35\eperiod
$$

\paragraph{Fractional covers and rounding.}
A \deffont{fractional cover} assigns a nonnegative weight $x_U$ to each $U\in\cV$, so that every witness contains subspaces of total weight at least one. Its cost is the sum of the subspace costs multiplied by their weights. We study the minimum fractional-cover cost at the intermediate dimension $e+h$:
\begin{align*}
\lambda=\min_{x_U\ge0}\quad
&\sum_{U\in\cV}x_Uq^{-(n-e-h)\dim U}\ecomma\\
\text{subject to}\quad
&\sum_{\substack{U\in\cV\\U\le A}}x_U\ge1
\qquad\text{for every }A\in\cA\eperiod
\numberthis\label{eq:primal-lp}
\end{align*}

The dual has a nonnegative weight $y_A$ for each witness. Each constraint limits the total weight of witnesses containing a fixed subspace:
\begin{align*}
\lambda=\max_{y_A\ge0}\quad
&\sum_{A\in\cA}y_A\ecomma\\
\text{subject to}\quad
&\sum_{\substack{A\in\cA\\U\le A}}y_A
\le q^{-(n-e-h)\dim U}
\qquad\text{for every }U\in\cV\eperiod
\numberthis\label{eq:dual-lp}
\end{align*}
Linear programming duality gives the common optimal value $\lambda$. Fix optimal solutions $x$ and $y$. We first round $x$ to obtain a lower bound on $\lambda$. We then use this lower bound and the dual weights $y_A$ in a second-moment argument to show that $C_k$ contains a witness with probability greater than $1/2$.

To lower-bound $\lambda$, round $x$ by retaining
$$
\cA':=\inset{U\in\cV:
x_U\ge\frac{q^{-h\dim U}}{\tau}}\eperiod
$$
If a witness $A$ contained no retained subspace, \cref{eq:witness-sum} would give
$$
\sum_{\{0\}<U\le A}x_U
<\frac1\tau\sum_{\{0\}<U\le A}q^{-h\dim U}
\le1\ecomma
$$
contradicting \cref{eq:primal-lp}. Thus, $\cA'$ is a cover. Its selection rule also gives
$$
\covercost_e(\cA')
\le\tau\sum_{U\in\cA'}x_Uq^{-(n-e-h)\dim U}
\le\tau\lambda\eperiod
$$
By the definition of $e$, every cover has cost at least $1/2$ at dimension $e$. Therefore,
$$
\lambda\ge\frac1{2\tau}\eperiod
$$

\paragraph{The weighted second moment.}
For a uniformly random $k$-dimensional code $C_k$, we use the dual weights to define
$$
X:=\sum_{A\in\cA}
y_Aq^{(n-k)\dim A}\mathbf{1}_{\{A\le C_k\}}\eperiod
$$
We first lower-bound its expectation. For a witness $A$ of dimension $d\le b$, we have
\begin{align*}
\frac{\PR{A\le C_k}}{q^{-(n-k)d}}
&=\prod_{i=0}^{d-1}
\frac{1-q^{i-k}}{1-q^{i-n}}\\
&\ge1-\sum_{i=0}^{d-1}q^{i-k}
\ge\frac{15}{16}\eperiod
\end{align*}
Indeed, $k\ge2b+2$, so the sum is at most
$(2^b-1)/2^{2b+2}\le1/16$. Hence
$$
\E{X}\ge\frac{15}{16}\lambda\eperiod
$$

To bound the second moment, we use the containment bound
$$
\PR{U\le C_k}\le q^{-(n-k)\dim U}
$$
for every fixed subspace $U$. A code contains both $A$ and $B$ exactly when it contains $A+B$, so
$$
\PR{A+B\le C_k}\le q^{-(n-k)\dim(A+B)}\eperiod
$$
Expanding $X^2$, applying this inequality, and then using
$\dim A+\dim B-\dim(A+B)=\dim(A\cap B)$ gives
\begin{align*}
\E{X^2}
&=\sum_{A,B\in\cA}
y_Ay_Bq^{(n-k)(\dim A+\dim B)}
\PR{A+B\le C_k}\\
&\le\sum_{A,B\in\cA}
y_Ay_Bq^{(n-k)(\dim A+\dim B-\dim(A+B))}\\
&=\sum_{A,B\in\cA}
y_Ay_Bq^{(n-k)\dim(A\cap B)}\eperiod
\end{align*}

We now use the dual constraints to bound the contribution of intersecting witnesses. These constraints control the total weight of witnesses containing each fixed subspace $U$. To apply them, we replace the term for $A\cap B$ by a sum over the nonzero subspaces shared by $A$ and $B$:
$$
q^{(n-k)\dim(A\cap B)}
\le1+\sum_{\{0\}<U\le A\cap B}q^{(n-k)\dim U}\eperiod
$$
Indeed, if the intersection is zero, both sides equal one. Otherwise, the sum includes $U=A\cap B$, whose term equals the left-hand side.

Substituting this inequality into our bound on $\E{X^2}$ gives
\begin{align*}
\E{X^2}
&\le\sum_{A,B\in\cA}y_Ay_B
\inparen{1+\sum_{\{0\}<U\le A\cap B}q^{(n-k)\dim U}}\\
&=\lambda^2+
\sum_{A\in\cA}y_A
\sum_{B\in\cA}y_B
\sum_{\{0\}<U\le A\cap B}q^{(n-k)\dim U}\eperiod
\end{align*}
The $\lambda^2$ term comes from summing $y_Ay_B\cdot1$ over all pairs:
$\sum_{A,B}y_Ay_B=(\sum_Ay_A)^2=\lambda^2$.

We now rearrange the remaining sum so that we can apply the dual constraints.
For each fixed $A$, choosing a witness $B$ and a nonzero subspace
$U\le A\cap B$ is equivalent to choosing a nonzero subspace $U\le A$
and a witness $B$ containing $U$. Exchanging the sums in this way gives
\begin{align*}
\E{X^2}
&\le\lambda^2+
\sum_{A\in\cA}y_A
\sum_{\{0\}<U\le A}q^{(n-k)\dim U}
\sum_{\substack{B\in\cA\\U\le B}}y_B\\
&\le\lambda^2+
\sum_{A\in\cA}y_A
\sum_{\{0\}<U\le A}q^{-h\dim U}\\
&\le\lambda\inparen{\lambda+\tau}\eperiod
\end{align*}
For the second inequality, we apply \cref{eq:dual-lp} to the innermost
sum and use $k=e+2h$. For the final inequality, we use the locality
bound \cref{eq:witness-sum} a second time.

Since $X>0$ implies $C_k\in\cP$, the Cauchy--Schwarz inequality and $\lambda\ge1/(2\tau)$ yield
\begin{align*}
\PR{C_k\in\cP}
&\ge\PR{X>0}
\ge\frac{\E{X}^2}{\E{X^2}}\\
&\ge\frac{\inparen{15/16}^2}{1+\tau/\lambda}
\ge\frac{\inparen{15/16}^2}{1+2\tau^2}
>\frac12\eperiod
\end{align*}
The final inequality uses $\tau<3/5$. Thus,
$$
\thresholddim(\cP)
\le k
=\expectationdim(\cP)+2b+2\eperiod
$$
\end{proof}

\section{From Common Projections to the Earlier Threshold Theorems}\label{sec:earlier-threshold-proofs}

We prove the threshold bounds announced in \cref{sec:intro-previous-thresholds}. We first give a condition under which common-projection covers locate the expectation threshold within \(O(b)\) dimensions. For row-symmetric properties, we obtain a sharper threshold formula and derive the original formula of~\cite{MRRSW20} as a consequence. For LCL properties, we recover the formula of~\cite{LMS25}. In both settings, we improve the quantitative bounds.

\subsection{Common Projections and Arbitrary Covers}\label{sec:projection-principle}

For a nonempty family $\cS$ of nonzero matrices with $n$ rows, let $\cP_{\cS}$ denote the property that a code contains the columns of some $M\in\cS$.\footnote{The covers depend only on the generated property: every generating family of witness subspaces contains all inclusion-minimal witnesses, and each of its members contains an inclusion-minimal witness.}

Suppose now that $\cS\subseteq\F_q^{n\times d}$ consists of rank-$d$ matrices. We retain the notation $\cS B$ and $\cA'_B$ from \cref{sec:intro-previous-thresholds}. To estimate the projection-expectation threshold using the projected-matrix counts, define
$$
\kappa(\cS):=
\max_{\substack{B\in\F_q^{d\times a}\\1\le a\le d,\ \rank B=a}}
\inparen{n-\frac{\log_q|\cS B|}{\rank B}}\eperiod
$$
Choose a rank-$a$ coefficient matrix $B$ attaining this maximum. The cover-cost bound from \cref{sec:intro-previous-thresholds} gives
$$
\covercost_k(\cA'_B)
\le|\cS B|q^{-(n-k)a}
=q^{-a(\kappa(\cS)-k)}\eperiod
$$
Since $a\ge1$, this cover has cost less than $1/2$ whenever $k<\kappa(\cS)-\log_q2$. Thus,
\begin{equation}\label{eq:ProjCovers}
\kappa(\cS)-\log_q2
\le\projectionexpectationdim(\cS)
\le\expectationdim(\cP_{\cS})\eperiod
\end{equation}

We next seek conditions on $\cS$ under which
$$
\expectationdim(\cP_{\cS})\le\kappa(\cS)+O(d)\eperiod
$$
Together with \cref{eq:ProjCovers}, this would place the projection-expectation threshold within $O(d)$ dimensions of the unrestricted expectation threshold. The following proposition gives an upper bound on the unrestricted expectation threshold from bounds on the probabilities of individual projection outcomes $MB=N$.

\begin{proposition}[Common projections and arbitrary covers]\label{prop:single-family-projection}
Let $\cP$ be a nontrivial increasing property of linear codes in $\F_q^n$, and let $M\in\F_q^{n\times d}$ be a random matrix, where $1\le d\le n$. Suppose that $\PR{\operatorname{col}(M)\in\cP}\ge1/2$ and that, for some real $r\ge0$,
$$
\PR{MB=N}\le q^{-(n-r)a}
$$
for every $1\le a\le d$, every rank-$a$ coefficient matrix $B\in\F_q^{d\times a}$, and every rank-$a$ matrix $N\in\F_q^{n\times a}$. Then
$$
\expectationdim(\cP)\le\lceil r\rceil+d\eperiod
$$
\end{proposition}

\begin{proof}
Put $k_0:=\lceil r\rceil+d$. If $k_0\ge n$, the conclusion is immediate, so assume $k_0<n$. We first show that
\begin{equation}\label{eq:UInCol(M)}
\PR{U\le\operatorname{col}(M)}
\le q^{-(n-k_0)\dim U}
\qquad\text{for every }U\le\F_q^n\eperiod
\end{equation}

Fix a subspace $U$ of dimension $1\le a\le d$, and choose a matrix $N_U\in\F_q^{n\times a}$ whose columns form a basis of $U$. The containment $U\le\operatorname{col}(M)$ holds exactly when $MB=N_U$ for some rank-$a$ coefficient matrix $B$. Indeed, preimages of the basis vectors are linearly independent because their images are. There are at most $q^{da}$ choices of $B$, so
\begin{align*}
\PR{U\le\operatorname{col}(M)}
&\le\sum_{\substack{B\in\F_q^{d\times a}\\\rank B=a}}
\PR{MB=N_U}\\
&\le q^{da}q^{-(n-r)a}\\
&\le q^{-(n-k_0)a}\eperiod
\numberthis\label{eq:projection-containment}
\end{align*}
For $\dim U>d$, containment is impossible, while for $U=\{0\}$ the bound holds with equality.

Let $\cA'$ be any cover of the minimal witnesses of $\cP$. Whenever $\operatorname{col}(M)\in\cP$, it contains a minimal witness and hence a member of $\cA'$. By \cref{eq:UInCol(M)} and the union bound,
$$
\frac12
\le\PR{\operatorname{col}(M)\in\cP}
\le\sum_{U\in\cA'}\PR{U\le\operatorname{col}(M)}
\le\covercost_{k_0}(\cA')\eperiod
$$
Every cover has cost at least $1/2$ at dimension $k_0$, so $\expectationdim(\cP)\le k_0$.
\end{proof}

For a projection-uniform family $\cS$, sample $M$ uniformly from $\cS$. Its column space always satisfies $\cP_{\cS}$, and $MB$ is uniform in $\cS B$ for every rank-$a$ coefficient matrix $B$. Hence, for every $N\in\F_q^{n\times a}$,
$$
\PR{MB=N}\le\frac1{|\cS B|}
\le q^{-(n-\kappa(\cS))a}\eperiod
$$
We apply \cref{prop:single-family-projection} with $\cP=\cP_{\cS}$ and $r=\kappa(\cS)$. Together with \cref{eq:ProjCovers}, this gives
$$
\projectionexpectationdim(\cS)
\le\expectationdim(\cP_{\cS})
\le\lceil\kappa(\cS)\rceil+d
\le\projectionexpectationdim(\cS)+d+1\eperiod
$$
The last inequality uses $\log_q2\le1$ and the integrality of $\projectionexpectationdim(\cS)$.

To pass from one matrix family to a property whose witnesses lie in several families, we take the union of their covers. The following lemma bounds the resulting loss in expectation threshold. We state it for families of witness subspaces.

\begin{lemma}[Decomposition of expectation thresholds]\label{lem:witness-decomposition}
Let $\cA=\bigcup_{j\in J}\cA_j$, where $J$ is finite and nonempty and each $\cA_j$ is a nonempty family of nonzero subspaces of $\F_q^n$. For each $j\in J$, let $\cA'_j$ be a cover of $\cA_j$. Then $\bigcup_{j\in J}\cA'_j$ is a cover of $\cA$ and
\begin{equation}\label{eq:witness-decomposition-cost}
\covercost_k\inparen{\bigcup_{j\in J}\cA'_j}
\le\sum_{j\in J}\covercost_k(\cA'_j)
\qquad(0\le k\le n)\eperiod
\end{equation}
Furthermore, the expectation thresholds satisfy
$$
\min_{j\in J}\expectationdim(\cP_{\cA_j})-\lceil\log_q|J|\rceil
\le\expectationdim(\cP_\cA)
\le\min_{j\in J}\expectationdim(\cP_{\cA_j})\eperiod
$$
\end{lemma}

\begin{proof}
Every witness belongs to some $\cA_j$ and therefore contains a member of $\cA'_j$. Thus, the union of the covers covers $\cA$. Its cost counts each distinct subspace once, proving \cref{eq:witness-decomposition-cost}.

Every cover of $\cA$ also covers each $\cA_j$, proving the upper bound on $\expectationdim(\cP_\cA)$. For the lower bound, put
$$
h:=\min_j\expectationdim(\cP_{\cA_j}),
\qquad
s:=\lceil\log_q|J|\rceil\eperiod
$$
If $h\le s$, the bound is immediate. Otherwise, for each $j$, choose a cover $\cA'_j$ of $\cA_j$ with cost less than $1/2$ at dimension $h-1$. These covers contain only nonzero subspaces, so decreasing the code dimension by $s$ reduces each member's contribution by a factor of at least $q^s$. By \cref{eq:witness-decomposition-cost},
$$
\covercost_{h-s-1}\inparen{\bigcup_j\cA'_j}
\le q^{-s}\sum_j\covercost_{h-1}(\cA'_j)
<\tfrac12|J|q^{-s}\le\tfrac12\eperiod
$$
Hence $\expectationdim(\cP_\cA)\ge h-s$.
\end{proof}

\subsection{Row-Symmetric Properties}\label{sec:row-symmetric-proof}

Let $\cP$ be a nontrivial increasing $b$-local property invariant under coordinate permutations. Represent each inclusion-minimal witness by all its ordered basis matrices, and group these matrices by their \deffont{row distribution}, the law of a uniformly chosen row. Let $\mathcal D$ be the set of these distributions, write $d_\tau\le b$ for the number of columns associated with $\tau\in\mathcal D$, and let $\mathcal T_n(\tau)$ contain all matrices with row distribution $\tau$. Every matrix in $\mathcal T_n(\tau)$ is a row permutation of one of the basis matrices, so its column space is again a minimal witness.

The cost of combining the families will depend on $|\mathcal D|$. For $d$ columns, a row distribution is specified by $q^d$ nonnegative integer multiplicities summing to $n$. Hence
\begin{equation}\label{eq:row-distribution-count}
|\mathcal D|
\le\sum_{d=1}^{b}\binom{n+q^d-1}{q^d-1}
\le b(n+1)^{q^b}\eperiod
\end{equation}
For fixed $q$ and $b$, the number of possible distributions with $b$ columns is a polynomial in $n$ of degree $q^b-1$. 

\paragraph{A threshold formula from exact matrix counting.}
For a row distribution $\sigma$ on $\F_q^a$, each multiplicity $n\sigma(x)$ is an integer. The number of matrices with these row multiplicities is
$$
M_n(\sigma):=|\mathcal T_n(\sigma)|
=\frac{n!}{\prod_{x\in\F_q^a}(n\sigma(x))!}\ecomma
$$
where $0!=1$.

For a rank-$a$ coefficient matrix $B\in\F_q^{d_\tau\times a}$, let $\tau B$ denote the law of $xB$ for $x\sim\tau$. Fix $M_0\in\mathcal T_n(\tau)$. Since column projections commute with row permutations, the matrices in $\mathcal T_n(\tau)B$ are exactly the row permutations of $M_0B$. Hence
$$
\mathcal T_n(\tau)B=\mathcal T_n(\tau B),
\qquad
|\mathcal T_n(\tau)B|=M_n(\tau B)\eperiod
$$
A row permutation taking one projected matrix to another also gives a bijection between their preimages in $\mathcal T_n(\tau)$. Thus, $\mathcal T_n(\tau)$ is projection-uniform.

We substitute these projected-matrix counts into the definition of $\kappa$ and put
\begin{align*}
K_\tau
&:=\kappa(\mathcal T_n(\tau))
=\max_{\substack{B\in\F_q^{d_\tau\times a}\\1\le a\le d_\tau,\ \rank B=a}}
\inparen{n-\frac{\log_q M_n(\tau B)}{\rank B}}\ecomma\\
\widehat\kappa_{\mathrm{row}}
&:=\min_{\tau\in\mathcal D}K_\tau\eperiod
\end{align*}
A code avoids $\cP$ exactly when it contains no witness from any of the row-distribution families. For each $\tau$, the maximizing projection gives a cover-cost bound below one when $k<K_\tau$. To make these bounds less than one for every family at the same dimension, we need $k<\widehat\kappa_{\mathrm{row}}$. We then sum the cover costs over the families, which can increase the bound by a factor of $|\mathcal D|$. The following corollary accounts for this sum through the term $\log_q(2|\mathcal D|)$.

\begin{corollary}[Row-symmetric threshold formula from exact matrix counting]\label{cor:row-multinomial-threshold}
With $\cP$, $\mathcal D$, and $\widehat\kappa_{\mathrm{row}}$ as above,
$$
\widehat\kappa_{\mathrm{row}}-\log_q(2|\mathcal D|)
\le\expectationdim(\cP)
\le\thresholddim(\cP)
\le\left\lceil\widehat\kappa_{\mathrm{row}}\right\rceil+3b+2\eperiod
$$
\end{corollary}

\begin{proof}
For each $\tau$, choose a projection $B_\tau$ attaining $K_\tau$, put $a_\tau:=\rank B_\tau$, and let $\cA'_\tau$ be the corresponding common-projection cover. For $k<\widehat\kappa_{\mathrm{row}}$, its cost is at most
$$
\covercost_k(\cA'_\tau)
\le M_n(\tau B_\tau)q^{-(n-k)a_\tau}
=q^{-a_\tau(K_\tau-k)}
\le q^{-(\widehat\kappa_{\mathrm{row}}-k)}\eperiod
$$
By \cref{lem:witness-decomposition}, the union of these covers covers all witnesses and has cost at most $|\mathcal D|q^{-(\widehat\kappa_{\mathrm{row}}-k)}$. This is less than $1/2$ whenever $k<\widehat\kappa_{\mathrm{row}}-\log_q(2|\mathcal D|)$, proving the lower bound.

For the upper bound, choose $\tau$ with $K_\tau=\widehat\kappa_{\mathrm{row}}$. Every cover of all witnesses also covers the witnesses represented by $\mathcal T_n(\tau)$. Since this family is projection-uniform, \cref{prop:single-family-projection} gives
$$
\expectationdim(\cP)
\le\expectationdim(\cP_{\mathcal T_n(\tau)})
\le\left\lceil\widehat\kappa_{\mathrm{row}}\right\rceil+d_\tau\eperiod
$$
Apply \cref{thm:intro-threshold} and use $d_\tau\le b$ to obtain the probability-threshold bound.
\end{proof}

For the upper bound, we need only one row-distribution family, so the common-projection comparison and \cref{thm:intro-threshold} give an $O(b)$ error. For the lower bound, we cover all families and incur $\log_q|\mathcal D|$ dimensions. Our general bound on this contribution is $q^b\log_q(n+1)+\log_q b$ (see \cref{eq:row-distribution-count}), so the lower error can remain much larger even with exact matrix counts.

\paragraph{The entropy formula.}
We recover the formula of~\cite{MRRSW20} by approximating the exact counts with entropy. For a distribution $\tau$ on $\F_q^d$, write
$$
H_q(\tau):=-\sum_x\tau(x)\log_q\tau(x)\ecomma
$$
with $0\log_q0=0$. Define
$$
\kappa_{\mathrm{row}}
:=n\min_{\tau\in\mathcal D}
\max_{\substack{B\in\F_q^{d_\tau\times a}\\1\le a\le d_\tau,\ \rank B=a}}
\inparen{1-\frac{H_q(\tau B)}{\rank B}}\ecomma
$$
and put $\Delta:=q^b\log_q(n+1)$. To pass from $\widehat\kappa_{\mathrm{row}}$ to $\kappa_{\mathrm{row}}$, we replace $\log_q M_n(\tau B)$ by $nH_q(\tau B)$ and keep the same families and projections.

For fixed $q$ and $b$, \cite[Theorem~2.8]{MRRSW20} locates the RLC threshold dimension at $\kappa_{\mathrm{row}}+o(n)$. We apply \cref{cor:row-multinomial-threshold} to obtain explicit bounds on the difference between $\kappa_{\mathrm{row}}$ and the threshold dimension, including when $q$ and $b$ grow with $n$.

\begin{corollary}[Row-symmetric threshold formula via entropy]\label{cor:intro-row-threshold}
With $\cP$, $\kappa_{\mathrm{row}}$, and $\Delta$ as above,
$$
\kappa_{\mathrm{row}}-\Delta-\log_q(2b)
\le\expectationdim(\cP)
\le\thresholddim(\cP)
\le\left\lceil\kappa_{\mathrm{row}}+\frac{\Delta}{b}\right\rceil+3b+2\eperiod
$$
\end{corollary}

\begin{proof}
For every row distribution $\sigma$ on $\F_q^a$, the standard multinomial bounds~\cite[Fact~2.3]{MRRSW20} give
$$
\frac{q^{nH_q(\sigma)}}{(n+1)^{q^a}}
\le M_n(\sigma)
\le q^{nH_q(\sigma)}\eperiod
$$
Applying these bounds to $\sigma=\tau B$ yields
$$
n\inparen{1-\frac{H_q(\tau B)}a}
\le n-\frac{\log_q M_n(\tau B)}a
\le n\inparen{1-\frac{H_q(\tau B)}a}
+\frac{q^a}{a}\log_q(n+1)\eperiod
$$
Since $q\ge2$, the sequence $q^a/a$ is nondecreasing for positive integers $a$. Thus, $q^a/a\le q^b/b$ whenever $1\le a\le b$. Maximizing over projections and minimizing over $\tau$ gives
\begin{equation}\label{eq:row-multinomial-entropy}
\kappa_{\mathrm{row}}
\le\widehat\kappa_{\mathrm{row}}
\le\kappa_{\mathrm{row}}+\frac{\Delta}{b}\eperiod
\end{equation}
Substitute \cref{eq:row-multinomial-entropy} and the bound $|\mathcal D|\le b(n+1)^{q^b}$ from \cref{eq:row-distribution-count} into \cref{cor:row-multinomial-threshold} to obtain the result.
\end{proof}

For the formula of~\cite{MRRSW20}, we obtain upper error $\frac{q^b}{b}\log_q(n+1)+O(b)$, compared with $O(q^{2b}\log_q(n+1))$ in~\cite[Lemma~2.7]{MRRSW20}. When we use $\widehat\kappa_{\mathrm{row}}$ instead, we avoid the entropy-approximation term and retain an upper error of $O(b)$.

\subsection{LCL Properties}\label{sec:lcl-proof}

In the LCL setting, we group witness matrices by their coordinate-wise constraints and the linear relations among their columns. Fix $b\ge2$. A \deffont{profile} $V=(V_1,\ldots,V_n)$ specifies subspaces $V_i\le\F_q^b$. A matrix $X\in\F_q^{n\times b}$ satisfies $V$ if its $i$th row lies in $V_i$ for every $i$. Given a nonempty family $\cF$ of profiles, let $\cP$ be the increasing property that a code contains the columns of a matrix with pairwise distinct columns satisfying some $V\in\cF$. We assume that $\cP$ is nontrivial. The $b$ columns certify membership, so $\cP$ is $b$-local.

The columns need not be linearly independent. We record their linear relations through the row space $U$ of $X$, since $\ker X=U^\perp$. To enforce distinctness, let $\mathcal U$ consist of the subspaces $U\le\F_q^b$ such that, for every $1\le r<s\le b$, some $u\in U$ has $u_r\ne u_s$. A matrix has pairwise distinct columns exactly when its row space belongs to $\mathcal U$.

Fix $V\in\cF$ and $U\in\mathcal U$, and consider the family of witness matrices satisfying $V$ whose row space is $U$. In each such matrix, row $i$ must lie both in $V_i$, as required by the profile, and in the row space $U$. Thus it lies in $V_i\cap U$.

To construct a common-projection cover for this family, we apply the same linear map on $U$ to every row of every matrix. If this map has kernel $W<U$, then each projected witness has dimension $\dim U-\dim W$. The image of $V_i\cap U$ under this map has dimension
$$
\dim(V_i\cap U)-\dim(V_i\cap W)\eperiod
$$
Each projected row lies in this image. Multiplying the sizes of these images therefore gives an upper bound on the number of projected matrices. Substituting this bound into the common-projection formula gives the expression of~\cite{LMS25}:
$$
\kappa_{\mathrm{lin}}
:=\min_{\substack{V\in\cF\\U\in\mathcal U}}
\max_{W<U}
\inparen{n-
\frac{\sum_{i=1}^n\inparen{\dim(V_i\cap U)-\dim(V_i\cap W)}}
{\dim U-\dim W}}\eperiod
$$
For each pair $(V,U)$, we maximize over projections, then minimize over the pairs. We use our cover characterization to bound the difference between $\kappa_{\mathrm{lin}}$ and the RLC threshold dimension, with explicit dependence on the locality and the number of profiles.

\begin{corollary}[LCL threshold formula]\label{cor:intro-lcl-threshold}
With $\cP$ and $\kappa_{\mathrm{lin}}$ as above,
$$
\kappa_{\mathrm{lin}}-b^2-\log_q(2|\cF|)
\le\expectationdim(\cP)
\le\thresholddim(\cP)
\le\left\lceil\kappa_{\mathrm{lin}}\right\rceil+3b+2\eperiod
$$
\end{corollary}

\begin{proof}
For each pair $(V,U)$, define the linear space
$$
\mathcal T_{V,U}:=
\inset{X\in\F_q^{n\times b}:
\text{row }i\text{ of }X\text{ lies in }V_i\cap U
\text{ for every }i}\eperiod
$$
The matrices in $\mathcal T_{V,U}$ with row space $U$ are the witnesses for this pair. For a fixed coefficient matrix $B$, we bound the number of distinct projected witness matrices by
$$
\left|\inset{XB:X\in\mathcal T_{V,U},\ \operatorname{row}(X)=U}\right|
\le|\mathcal T_{V,U}B|\eperiod
$$

Fix $1\le a\le\dim U$, and choose a coefficient matrix $B\in\F_q^{b\times a}$ such that every vector in $\F_q^a$ equals $uB$ for some $u\in U$. Put
$$
W:=\inset{u\in U:uB=0}\eperiod
$$
Then $\dim U-\dim W=a$. For $X\in\mathcal T_{V,U}$, row $i$ of $XB$ can be any vector in $(V_i\cap U)B$. The map $u\mapsto uB$ on $V_i\cap U$ has kernel $V_i\cap W$, so there are exactly
$$
q^{\dim(V_i\cap U)-\dim(V_i\cap W)}
$$
possible values for this row. Every combination of these row values gives a matrix in $\mathcal T_{V,U}B$: we can choose a preimage of each row independently in $V_i\cap U$. Multiplying the numbers of choices gives
\begin{equation}\label{eq:lcl-projection-count}
|\mathcal T_{V,U}B|
=q^{\sum_{i=1}^n\inparen{\dim(V_i\cap U)-\dim(V_i\cap W)}}\eperiod
\end{equation}

We can realize every proper subspace $W<U$ in the definition of $\kappa_{\mathrm{lin}}$ as the kernel of a projection on $U$. Given $W$, put $a:=\dim U-\dim W$ and choose a surjective linear map $\varphi:U\to\F_q^a$ with kernel $W$. Extend $\varphi$ linearly to $\F_q^b$, and let $B\in\F_q^{b\times a}$ represent this extension. Then $uB=\varphi(u)$ for every $u\in U$, so $B$ maps $U$ onto $\F_q^a$ and has kernel $W$ on $U$.

For the lower bound, fix a pair $(V,U)$ and choose $W<U$ attaining the inner maximum in the definition of $\kappa_{\mathrm{lin}}$. Choose a coefficient matrix $B$ realizing this kernel on $U$, with $a:=\dim U-\dim W$ columns. Since $\kappa_{\mathrm{lin}}$ is the minimum of these inner maxima over all pairs, \cref{eq:lcl-projection-count} gives
$$
n-\frac{\log_q|\mathcal T_{V,U}B|}{a}
\ge\kappa_{\mathrm{lin}}\eperiod
$$ 
If $X$ has row space $U$, then $XB$ has rank $a$ and $\operatorname{col}(XB)\le\operatorname{col}(X)$. The projected column spaces therefore cover this pair's witnesses, at cost at most
$$
|\mathcal T_{V,U}B|q^{-(n-k)a}
\le q^{-a(\kappa_{\mathrm{lin}}-k)}
\le q^{-(\kappa_{\mathrm{lin}}-k)}
\qquad(k<\kappa_{\mathrm{lin}})\eperiod
$$
There are at most $q^{b^2}$ choices of $U$: each subspace has a reduced row-echelon basis that we can pad to a $b\times b$ matrix. Pairs with no witnesses contribute no cover members. By \cref{lem:witness-decomposition}, the union of the covers has cost at most
$$
|\cF|q^{b^2}q^{-(\kappa_{\mathrm{lin}}-k)}\eperiod
$$
This is less than $1/2$ whenever $k<\kappa_{\mathrm{lin}}-b^2-\log_q(2|\cF|)$, proving the lower bound.

For the upper bound, choose a pair attaining $\kappa_{\mathrm{lin}}$, write $\mathcal T:=\mathcal T_{V,U}$, and put $k_0:=\lceil\kappa_{\mathrm{lin}}\rceil+b$. If $k_0\ge n$, the bound $\expectationdim(\cP)\le k_0$ is immediate. Otherwise, sample $X$ uniformly from $\mathcal T$. We verify the two hypotheses of \cref{prop:single-family-projection} with $M=X$, $d=b$, and $r=\kappa_{\mathrm{lin}}$.

First fix $1\le a\le b$, a rank-$a$ coefficient matrix $B\in\F_q^{b\times a}$, and a rank-$a$ matrix $N\in\F_q^{n\times a}$. If the restriction of $u\mapsto uB$ to $U$ has rank less than $a$, then $XB=N$ is impossible. Otherwise, every nonempty fiber of the linear map $X\mapsto XB$ on $\mathcal T$ has the same size. For the chosen pair $(V,U)$, the inner maximum in the definition of $\kappa_{\mathrm{lin}}$ equals $\kappa_{\mathrm{lin}}$. Applying \cref{eq:lcl-projection-count} to the kernel of $u\mapsto uB$ on $U$ therefore gives
$$
n-\frac{\log_q|\mathcal T B|}{a}
\le\kappa_{\mathrm{lin}}\eperiod
$$
Hence $|\mathcal T B|\ge q^{(n-\kappa_{\mathrm{lin}})a}$, and
$$
\PR{XB=N}\le\frac1{|\mathcal T B|}
\le q^{-(n-\kappa_{\mathrm{lin}})a}\eperiod
$$

Next we show that the rows of $X$ span $U$ with probability greater than $1/2$, so that $X$ is a witness for this pair. If the rows do not span $U$, some nonzero linear functional on $U$ vanishes on every row. For each such functional, \cref{eq:lcl-projection-count} with $a=1$ and the equal-sized fibers bound the probability of this zero outcome by $q^{-(n-\kappa_{\mathrm{lin}})}$. Functionals that differ by a nonzero scalar have the same kernel, so
$$
\PR{\operatorname{row}(X)\ne U}
\le\frac{q^{\dim U}-1}{q-1}q^{-(n-\kappa_{\mathrm{lin}})}
<\frac12\ecomma
$$
where we used $n-\kappa_{\mathrm{lin}}\ge b+1\ge\dim U+1$, which follows from $k_0<n$. Thus $\PR{\operatorname{col}(X)\in\cP}>1/2$, and \cref{prop:single-family-projection} gives $\expectationdim(\cP)\le k_0$. Applying \cref{thm:intro-threshold} proves the upper bound.
\end{proof}

When $b^2+\log_q|\cF|=o(n)$, our bounds recover the threshold rate $\kappa_{\mathrm{lin}}/n+o(1)$ of~\cite[Eq.~(14) and Theorem~4.4]{LMS25}. We improve the upper error from $b^2+O(1)$ to $O(b)$ dimensions. The lower-bound terms $b^2$ and $\log_q|\cF|$ still account for combining the candidate row spaces and profiles, respectively. When the profile-count bound is independent of $q$, increasing the alphabet size reduces the contribution $\log_q|\cF|$.
\section*{AI Disclosure}
We used GPT-6 Astra to assist with the write-up of the proofs in Section 3 and the Appendix, working from our proof sketches. We also used it for general editing, proofreading and error finding. The ideas, approach and technical analysis are all due to the authors,  as is the presentation apart from the assistance described above. The authors take full responsibility for all content.

\section*{Human Disclosure}
We thank Sepehr Assadi and Sanjeev Khanna for acting as matchmakers and bringing us together for this collaboration.

\clearpage
\printbibliography

\clearpage
\appendix
\crefname{appendix}{Appendix}{Appendices}
\Crefname{appendix}{Appendix}{Appendices}

\section{Proof of \cref{cor:intro-transfer}: Transfer of Local Guarantees}\label[appendix]{app:transfer-proof}
\begin{proof}
Put $d:=\thresholddim\inparen{\cP}-2b-4$. The rate assumption gives $d\ge nR+t\ge0$, while \cref{thm:intro-threshold} gives
$$
d<\expectationdim\inparen{\cP}\eperiod
$$
Choose a family $\cA$ of witnesses to failure of $\cP$, each of dimension at most $b$. By the definition of the expectation threshold, $\cA$ has a cover $\cA'$ with $\covercost_d\inparen{\cA'}<1/2$. Discard each cover member that lies in no witness. Every remaining member has dimension at most $b$, and the cost bound excludes the zero subspace.

Let $G$ be the event in the definition of conditional pseudorandomness. If $C$ fails $\cP$, it contains some member of $\cA'$. Consequently,
\begin{align*}
\PR{C\notin\cP}
&\le\PR{G^c}+\sum_{U\in\cA'}\PR{U\le C\text{ and }G}\\
&\le\delta+\sum_{U\in\cA'}q^{-n(1-R)\dim U}\\
&\le\delta+q^{-t}\sum_{U\in\cA'}q^{-(n-d)\dim U}\\
&<\delta+\tfrac12q^{-t}\eperiod
\end{align*}
The third inequality uses $d-nR\ge t$ and $\dim U\ge1$.
\end{proof}

\section{\Cref{cor:intro-ensembles}: Pseudorandom Structured Ensembles}\label[appendix]{app:ensemble-guarantees}\label[appendix]{app:ensemble-pseudorandomness}
\begin{corollary}[Pseudorandom structured ensembles]\label{cor:intro-ensembles}
Let $b\ge1$ be an integer, which may depend on $n$ except in Part~\ref{item:ensembles-gallager}. The following ensembles satisfy the pseudorandomness hypotheses of \cref{cor:intro-transfer}.
\begin{enumerate}
\item\label{item:ensembles-low-bias} \emph{Randomly punctured low-bias codes.}
Let $D\le\F_q^N$ have dimension $k\le n$, and write $R_0=k/n$. Sample $n$ coordinates independently and uniformly from $[N]$, with replacement, and restrict every codeword of $D$ to the sampled coordinates. Suppose that $D$ has bias at most $\beta$.\footnote{This means that $|N^{-1}\sum_{i=1}^N\chi(c_i)|\le\beta$ for every nonzero $c\in D$ and every nontrivial additive character $\chi$ of $\F_q$.} For every $\eta>0$ with $R_0+\eta\le1$, the resulting ensemble is \pseudorandom{b}{R_0+\eta} provided
$$
\beta\le\frac{q^\eta-1}{q^b}\eperiod
$$
The same conclusion holds for the expander-walk puncturing of~\cite{PP23b}: sample the coordinates along a random walk on an undirected regular graph on $[N]$, starting at a uniformly chosen vertex. If its normalized adjacency matrix has operator norm at most $\lambda$ on the orthogonal complement of the constant vector, it suffices that
$$
\beta+\lambda\le\frac{q^\eta-1}{q^b}\eperiod
$$
In either sampling model, the punctured code has dimension $k$ except with probability at most $q^{-n(1-R_0-\eta)}/(q-1)$.

\item\label{item:ensembles-linearized} \emph{Codes from linearized polynomials.}
Let $0\le k\le n$ and put $R_0=k/n$. The linearized-polynomial ensemble of~\cite{MRSY25}, defined in \cref{app:linearized-polynomials}, is \pseudorandom{b}{R_0}. Its dimension is $k$ except with probability at most $\frac{q^k-1}{q-1}q^{-n}$.

\item\label{item:ensembles-gallager} \emph{Gallager's random LDPC codes.}
Fix $q$ and $0<R_{\max}<1$. Consider Gallager's ensemble, defined in \cref{app:gallager-pseudorandomness}, with odd check degree $s$ and design rate $0\le R_0\le R_{\max}$. We restrict to admissible parameters $s\mid n$ and $(1-R_0)s\in\N$. Every sampled code has dimension at least $R_0n$. For fixed $b$ and $\eta>0$ with $R_{\max}+\eta<1$, every sufficiently large fixed $s$ gives an \almostpseudorandom{b}{R_0+\eta}{o(1)} ensemble.

\item\label{item:ensembles-small-bias-matrices}
\emph{Kernels of small-bias matrices.}
Let $1\le r\le n$, and sample an $\varepsilon$-biased matrix
$A\in\F_q^{n\times r}$ with $\varepsilon\le q^{-rb}$.
Then $C:=\ker A^{\mathsf T}$ comes from a
\pseudorandom{b}{1-(r-1)/n} ensemble, and every sampled code has dimension at least $n-r$.
An explicit distribution of $A$ uses
$O(\log_2(nr)+rb\log_2 q)$ random bits.

\end{enumerate}
\end{corollary}

\subsection{Proof of Part~\ref{item:ensembles-low-bias}: Low-Bias Punctures}\label[appendix]{app:low-bias-punctures}

We include both independent coordinate sampling and the expander-walk sampling of~\cite{PP23b}. The latter visits coordinates along a random walk rather than choosing each coordinate independently. To analyze the resulting code, we use the following form of the expander hitting bound, retaining the endpoint factors.

\begin{sloppypar}
Let $i_1,\ldots,i_\ell$ be the vertices of a random walk on an undirected regular graph on $[N]$, starting uniformly. Write $P$ for its normalized adjacency matrix and $J$ for the orthogonal projection onto the constant vectors, and suppose that $\lVert P-J\rVert_2\le\lambda$. For sets $T_1,\ldots,T_\ell\subseteq[N]$ of densities $p_j:=|T_j|/N\le p$, we have
\begin{align*}
\PR{i_j\in T_j\text{ for every }1\le j\le\ell}
&\le p(p+\lambda)^{\ell-1}
\le(p+\lambda)^\ell\eperiod
\numberthis\label{eq:expander-hitting}
\end{align*}
To see this, let $D_j$ be the diagonal projection onto the coordinates in $T_j$. The decomposition $P=J+(P-J)$ gives
$$
\lVert D_jPD_{j+1}\rVert_2
\le\sqrt{p_jp_{j+1}}+\lambda\eperiod
$$
The probability in \cref{eq:expander-hitting} equals $e^{\mathsf T}D_1PD_2\cdots PD_\ell e$, where $e:=N^{-1/2}(1,\ldots,1)^{\mathsf T}$. Since $\lVert D_1e\rVert_2,\lVert D_\ell e\rVert_2\le\sqrt p$, multiplying the bounds proves \cref{eq:expander-hitting}. The sets may differ between steps, which lets us prescribe a different joint symbol pattern at every sampled coordinate.
\end{sloppypar}

\begin{proof}
Fix a subspace $U\le\F_q^n$ of dimension $1\le d\le b$ and a basis of $U$. If the punctured code contains $U$, some linearly independent $d$-tuple in $D$ maps to that basis. Fix such a tuple. For each output position, the required joint symbol pattern defines a set of mother-code coordinates. Lemma~6.4(1) of~\cite{GM22} bounds the density of each set by $q^{-d}+\beta$. Independent sampling gives probability at most $(q^{-d}+\beta)^n$ of matching all the required patterns. \cref{eq:expander-hitting} gives at most $(q^{-d}+\beta+\lambda)^n$ for expander-walk sampling.

Set $\lambda=0$ in the independent case. Since $D$ contains at most $q^{kd}$ candidate tuples, both sampling rules satisfy
\begin{align*}
\PR{U\le C}
&\le q^{kd}\inparen{q^{-d}+\beta+\lambda}^n\\
&=q^{-n(1-R_0)d}\inparen{1+q^d(\beta+\lambda)}^n\\
&\le q^{-n(1-R_0)d+n\eta}\\
&\le q^{-n(1-R_0-\eta)d}\eperiod
\end{align*}
The hypothesis gives $1+q^d(\beta+\lambda)\le q^\eta$, and the last inequality uses $d\ge1$. This proves the claimed pseudorandomness for both sampling rules, without a symmetry assumption on the property or a fixed-locality assumption.

To bound dimension loss, observe that puncturing fails to be injective on $D$ exactly when it maps some nonzero word to zero. Each nonzero word vanishes on at most a $q^{-1}+\beta$ fraction of the mother-code coordinates. Independence, or \cref{eq:expander-hitting} for the walk, bounds the probability that it punctures to zero by $(q^{-1}+\beta+\lambda)^n$. Scalar multiples have the same zero event, so a union bound over the one-dimensional subspaces of $D$ gives
$$
\PR{\dim C<k}
\le\frac{q^k-1}{q-1}\inparen{q^{-1}+\beta+\lambda}^n
\le\frac{q^{-n(1-R_0-\eta)}}{q-1}\eperiod
$$
Here, $\lambda=0$ for independent sampling, and the last inequality follows from $b\ge1$ and $1+q(\beta+\lambda)\le q^\eta$.
\end{proof}

\paragraph{Randomness used by the walk.}
Sampling $n$ coordinates along a walk on a regular graph of degree $\Delta$ uses $O(\log N+n\log\Delta)$ random bits, compared with $O(n\log N)$ for independent sampling.\footnote{For arbitrary $N$ and $\Delta$, we count the expected number of random bits needed for exact uniform sampling.} By taking a suitable power of a constant-degree expander, allowing parallel edges and self-loops, we can achieve spectral parameter at most $\lambda$ with $\log\Delta=O(1+\log(1/\lambda))$~\cite{PP23b}.

For fixed $q$ and $0<\eta\le1$, suppose that the mother code has bias
$$
\beta\le\frac{q^\eta-1}{2q^b}\eperiod
$$
We can choose the sampling graph with spectral parameter at most $(q^\eta-1)/(2q^b)$ and $\log\Delta=O(b+\log(1/\eta))$. The puncturing then meets the pseudorandomness condition in \cref{cor:intro-ensembles}, and sampling uses
$$
O\inparen{\log N+n\inparen{b+\log(1/\eta)}}
$$
random bits. For fixed $b$ and $\eta$, this is $O(n)$ whenever $\log N=O(n)$.

\subsection{Proof of Part~\ref{item:ensembles-linearized}: Linearized-Polynomial Codes}\label[appendix]{app:linearized-polynomials}

We use the following ensemble from~\cite{MRSY25}. Fix $0\le k\le n$ and a $k$-dimensional $\F_q$-subspace $V\le\F_{q^n}$, and identify $\F_{q^n}$ with $\F_q^n$ through a fixed basis. Sample $a_0,\ldots,a_{b-1}$ independently and uniformly from $\F_{q^n}$, and define
$$
f(x):=\sum_{i=0}^{b-1}a_i x^{q^i}\eperiod
$$
The sampled code is $f(V)$, and its design rate is $R_0=k/n$.
\begin{proof}
Fix a subspace $U\le\F_q^n$ of dimension $1\le d\le b$ and an ordered basis of $U$. If $f(V)$ contains $U$, some linearly independent $d$-tuple in $V$ maps to this basis. Lemma~3.1 of~\cite{MRSY25} states that the values of $f$ on any such tuple are independent and uniform in $\F_{q^n}$. There are at most $q^{kd}$ candidate tuples, so
$$
\PR{U\le f(V)}\le q^{kd}q^{-nd}=q^{-n(1-R_0)d}\eperiod
$$
For every fixed nonzero $x\in V$, the value $f(x)$ is uniform in $\F_{q^n}$: conditional on $a_1,\ldots,a_{b-1}$, the term $a_0x$ is uniform. Since $f$ is $\F_q$-linear, failure of injectivity on $V$ means that its kernel contains a one-dimensional subspace of $V$. A union bound over these subspaces gives
$$
\PR{\dim f(V)<k}
\le\frac{q^k-1}{q-1}q^{-n}\eperiod
$$
Sampling the $b$ coefficients uses $O(bn\log q)$ random bits.
\end{proof}

\subsection{Proof of Part~\ref{item:ensembles-gallager}: Gallager Codes}\label[appendix]{app:gallager-pseudorandomness}

We use Gallager's ensemble from~\cite{MRRSW20}. For odd $s$ with $s\mid n$ and $(1-R_0)s\in\N$, we stack $(1-R_0)s$ independent layers. Each layer partitions the $n$ coordinates uniformly at random into blocks of size $s$ and imposes one parity check on each block, with independent uniform coefficients in $\F_q\setminus\{0\}$. The resulting parity-check matrix has $(1-R_0)n$ rows, giving design rate $R_0$. Its rows may be dependent, so its kernel always has dimension at least $R_0n$.

For fixed locality, choose a constant $\alpha>0$ small enough that Theorem~2.14 of~\cite{MRRSW20} guarantees minimum distance at least $\alpha n$ with probability $1-o(1)$ for every design rate at most $R_{\max}$. Let $G$ denote this minimum-distance event. A subspace containing a nonzero vector of weight less than $\alpha n$ cannot lie in $C$ on $G$. For every other subspace $U$ of dimension at most $b$, Lemma~2.13 of~\cite{MRRSW20}, applied with its error parameter equal to $\eta$, gives
$$
\PR{U\le C\text{ and }G}
\le q^{-n(1-\eta)(1-R_0)\dim U}
\le q^{-n(1-R_0-\eta)\dim U}\eperiod
$$
Since $q,b,R_{\max},\eta$ are fixed, one sufficiently large odd check degree makes both estimates hold.

\subsection{Proof of Part~\ref{item:ensembles-small-bias-matrices}: Kernels of Small-Bias Matrices}\label[appendix]{app:small-bias-matrices}

\paragraph{Containment probabilities.}
Let $A\in\F_q^{n\times r}$ be an $\varepsilon$-biased random matrix, and fix a subspace $U\le\F_q^n$ of dimension $d$. Choose a matrix $X\in\F_q^{n\times d}$ whose columns form a basis of $U$. The event $U\le\ker A^{\mathsf T}$ is exactly $A^{\mathsf T}X=0$. Since $X$ has rank $d$, the linear map $A\mapsto A^{\mathsf T}X$ is surjective onto $\F_q^{r\times d}$.

Fix a nontrivial additive character $\chi$ of $\F_q$. Character orthogonality gives
$$
\PR{A^{\mathsf T}X=0}
=q^{-rd}\sum_{Z\in\F_q^{r\times d}}
\E{\chi\inparen{\sum_{i=1}^r\sum_{j=1}^d Z_{ij}(A^{\mathsf T}X)_{ij}}}\eperiod
$$
The term $Z=0$ contributes $q^{-rd}$. For every $Z\ne0$, surjectivity ensures that the expression inside $\chi$ is a nonzero linear form in the entries of $A$. Small bias therefore bounds the corresponding character expectation in absolute value by $\varepsilon$. Summing the remaining terms gives
$$
\left|\PR{U\le\ker A^{\mathsf T}}-q^{-rd}\right|
\le(1-q^{-rd})\varepsilon\eperiod
$$

Under the hypothesis $\varepsilon\le q^{-rb}$, every $U$ of dimension $1\le d\le b$ consequently satisfies
$$
\PR{U\le\ker A^{\mathsf T}}
\le q^{-rd}+q^{-rb}
\le2q^{-rd}
\le q^{-(r-1)d}\eperiod
$$
Thus the kernel ensemble is \pseudorandom{b}{1-(r-1)/n}. Rank--nullity also gives $\dim\ker A^{\mathsf T}\ge n-r$ for every sampled matrix.

Using standard small-bias constructions over finite fields~\cite{AGHP90}, we can sample an $\varepsilon$-biased vector in $\F_q^{nr}$ using
$$
O\inparen{\log_2(nr)+\log_2 q+\log_2(1/\varepsilon)}
$$
random bits. Arranging its coordinates into an $n\times r$ matrix and taking $\varepsilon=q^{-rb}$ gives the claimed $O(\log_2(nr)+rb\log_2 q)$ randomness bound.

\section{Proof of \cref{cor:intro-binary-list-decoding}: Binary List Decoding}\label[appendix]{app:binary-list-decoding}\label[appendix]{app:puncturing-applications}

\paragraph{Expander-walk variant.}
The expander-walk variant of \cref{cor:intro-binary-list-decoding} samples $n$ coordinates along a random walk on an undirected regular graph on $[N]$, starting uniformly. Let $\lambda$ be the graph's spectral parameter as defined in \cref{cor:intro-ensembles}. The same conclusion holds if the mother code has bias at most $\beta$ and
$$
\beta+\lambda\le\frac{2^{-L}}{2n}\eperiod
$$
We prove the independent and expander-walk versions together.

\begin{proof}
Let $C$ be the punctured code. We bound the two possible failures separately: list-decoding failure and dimension loss will each have probability at most $2^{-t-2}$. Adding these bounds gives $\tfrac12 2^{-t}$, the total failure probability allowed in the corollary.

For list decoding, subtracting one codeword from an offending $(L+1)$-tuple shows that $\cP$ is $L$-local. Write $\beta$ for the bias bound of $D$, and let $\lambda$ be the graph's spectral parameter for expander-walk sampling, or zero for independent sampling. In both cases, the hypothesis gives $\beta+\lambda\le 2^{-L}/(2n)$. Since
$$
1+2^L(\beta+\lambda)\le 2^{1/n}\ecomma
$$
\cref{cor:intro-ensembles} makes the puncturing ensemble \pseudorandom{L}{(k+1)/n}. The dimension assumption gives
$$
k+1\le\thresholddim\inparen{\cP}-2L-4-(t+1)\eperiod
$$
Applying \cref{cor:intro-transfer} with slack $t+1$ therefore yields
$$
\PR{C\notin\cP}\le 2^{-t-2}\eperiod
$$

For dimension loss, Part~\ref{item:ensembles-low-bias} of \cref{cor:intro-ensembles}, applied with $\eta=1/n$, gives
$$
\PR{\dim C<k}
\le 2^{k-n+1}
\le 2^{-2L-5-t}
\le 2^{-t-2}\eperiod
$$
Here we used the dimension assumption
$k\le\thresholddim\inparen{\cP}-2L-6-t\le n-2L-6-t$.
A union bound over the two failures now gives
$$
\PR{\dim C=k\text{ and }C\in\cP}
\ge1-2^{-t-2}-2^{-t-2}
=1-\tfrac12 2^{-t}\eperiod
$$
\end{proof}

\section{Proof of \cref{cor:intro-hashing}: Hashing from Small-Bias Matrices}\label[appendix]{app:hashing}
\begin{proof}[Proof of \cref{cor:intro-hashing}]
We first prove the maximum-load guarantee, which is immediate if $N\le L$. Assume that $N>L$, and let $\cP$ denote $(S,L)$-list-decodability for linear codes in $\F_q^n$. This property is nontrivial, decreasing, and $L$-local.

To apply the reduction, we first show that a uniformly random surjective linear map $g:\F_q^n\to\F_q^m$ has maximum load at most $L$ on $S$ with probability $1-o(1)$. Put
$$
\Delta:=m-\log_qN\ge0\eperiod
$$
Choose $c_q\ge\max\{16q,32q/\ln q\}$. For all sufficiently large $m$, the definition of $L$ gives
$$
L\ge8q,
\qquad
\frac{8qm}{L}
\le\frac{\ln q}{4}\inparen{\Delta+\log_qm}\eperiod
$$
The maximum-load bound in~\cite[Theorem~3.2]{KM26} therefore gives
\begin{align*}
\PR{\max_{y\in\F_q^m}|S\cap g^{-1}(y)|>L}
&\le \frac{8N}{q^mL}
e^{8q(m+2\lceil\log_qL\rceil)/L}\\
&\le O_q\inparen{\frac{m^{1/4}}{L}q^{-3\Delta/4}}\\
&\le O_q\inparen{
\frac{\Delta+\log_qm}{m^{3/4}}q^{-3\Delta/4}}\\
&=O_q\inparen{\frac{\log m}{m^{3/4}}}
=o(1)\eperiod
\end{align*}
For the second inequality, we use the bound on $8qm/L$ and the fact that $\lceil\log_qL\rceil/L$ is bounded. For the final estimate, we use $\Delta q^{-3\Delta/4}=O_q(1)$. Thus the probability bound is uniform over $1\le N\le q^m$.

Since $\ker g$ is an RLC of dimension $n-m$, we obtain
$$
\thresholddim(\cP)>n-m
$$
for all sufficiently large $m$.

Put $C:=\ker A^{\mathsf T}$. By \cref{cor:intro-ensembles}, the ensemble of $C$ is \pseudorandom{L}{R}, where
$$
R:=1-\frac{r-1}{n}\eperiod
$$
Our choice of $r$ gives
$$
nR=n-m-2L-4-t
\le\thresholddim(\cP)-2L-4-t\eperiod
$$
Applying \cref{cor:intro-transfer} yields
$$
\PR{C\notin\cP}\le\tfrac12q^{-t}\eperiod
$$
The hashing identity from \cref{sec:intro-properties} identifies this failure event with maximum load greater than $L$ on $S$.

The sampling bound from \cref{app:small-bias-matrices}, with $b=L$, gives the claimed number of random bits. Finally, since $N\le q^m$, we have
$$
L\le\left\lceil\frac{c_qm}{\log_qm}\right\rceil
=O_q(m/\log m)=o(m)\eperiod
$$
Thus, if $t=o(m)$, then
$$
r=m+o(m),
\qquad
q^r=q^{m+o(m)}\eperiod
$$
Choosing also $t\to\infty$ makes the failure probability vanish.
\end{proof}

\section{Verification of \cref{ex:intro-common-projection}: Common Projections versus Arbitrary Covers}\label[appendix]{app:common-projection-example}

We verify the construction and threshold values in \cref{ex:intro-common-projection}. Throughout, $n\ge8$ is divisible by four and $0\ne v\in\F_2^n$.

\paragraph{Constructing the witness family.}
Identify $\F_2^n$ with $\F_4^{n/2}$. The one-dimensional $\F_4$-subspaces give $(2^n-1)/3$ binary planes that intersect pairwise only at zero. Exclude the plane containing $v$ and choose $2^{n/2}$ of the remaining planes to form $\cD$. This is possible since
$$
\frac{2^n-4}{3}\ge2^{n/2}\eperiod
$$
Set
$$
\cA:=\inset{A\le\F_2^n:\dim A=2,\ v\in A}\cup\cD\eperiod
$$
For each witness in $\cA$, choose one or more ordered bases, and let $\cS\subseteq\F_2^{n\times2}$ consist of the resulting matrices.

\paragraph{The unrestricted expectation threshold.}
The cover $\inset{\langle v\rangle}\cup\cD$ has cost
$$
2^{k-n}+2^{n/2}2^{2(k-n)}\eperiod
$$
At $k=3n/4-1$, this cost equals $2^{-n/4-1}+1/4<1/2$, so
$$
\expectationdim(\cP_\cA)\ge\frac{3n}{4}\eperiod
$$
Conversely, any cover that excludes zero must contain a distinct nonzero subspace inside each member of $\cD$. These subspaces have dimension at most two, so at $k=3n/4$ their contributions total at least
$$
|\cD|2^{2(k-n)}=1\eperiod
$$
A cover containing zero also costs at least one. Thus
$$
\expectationdim(\cP_\cA)=\frac{3n}{4}\eperiod
$$

\paragraph{The projection-expectation threshold.}
Fix a common projection $B\in\F_2^{2\times a}$ of rank $a\in\inset{1,2}$. If $a=1$, the projected cover contains a line from every plane in $\cD$. These lines are distinct, regardless of the representing bases. The cover therefore has cost at least $2^{n/2}2^{k-n}$, which reaches $1/2$ at $k=n/2-1$. No rank-one projection can have a threshold greater than $n/2-1$.

If $a=2$, the projection preserves every witness plane, so the projected cover is exactly $\cA$. There are $2^{n-1}-1$ planes containing $v$, corresponding to the lines of $\F_2^n/\langle v\rangle$. Hence this cover has cost
$$
\inparen{2^{n-1}-1+2^{n/2}}2^{2(k-n)}\eperiod
$$
At $k=n/2$, the cost is $1/2+2^{-n/2}-2^{-n}>1/2$. At $k=n/2-1$, it is one quarter of this quantity and is less than $1/2$. Thus the rank-two cover has threshold exactly $n/2$. Taking the better of the two projection ranks gives
$$
\projectionexpectationdim(\cS)=\frac n2
$$
for every choice of representing matrices.

\paragraph{The probability threshold.}
Since every witness has dimension two, \cref{thm:intro-threshold} gives
$$
\frac{3n}{4}
\le\thresholddim(\cP_\cA)
\le\frac{3n}{4}+6\eperiod
$$
This proves the claimed constant error around $3n/4$ and the linear gap from the projection-expectation threshold.

\section{A Binary Example with a Logarithmic Threshold Gap}\label[appendix]{app:full-support-threshold-gap}

Let $b\ge1$ be an integer, put $n:=2^b-1$, and let $\cP_b$ be the property of having full coordinate support: a code $C\le\F_2^n$ satisfies $\cP_b$ if, for every coordinate $i$, some $c\in C$ has $c_i=1$. This property is increasing and invariant under coordinate permutations. We prove that its locality is exactly $b$ and that, for all sufficiently large $b$,
$$
\thresholddim(\cP_b)=b+1,
\qquad
\expectationdim(\cP_b)=b-\log_2 b+O(1)\eperiod
$$
In particular,
$$
\thresholddim(\cP_b)-\expectationdim(\cP_b)
=\log_2 b+O(1)\eperiod
$$

\begin{proof}
We first determine the locality and the probability threshold, then bound the expectation threshold from both sides.

\paragraph{Locality.}
Let $A$ be an inclusion-minimal member of $\cP_b$, and put $r:=\dim A$. Each coordinate restricts to a nonzero linear functional $\ell_i:A\to\F_2$. Every hyperplane $H<A$ fails to have full support, so some coordinate vanishes on $H$. The corresponding functional $\ell_i$ is nonzero on $A$, and therefore $\ker\ell_i=H$. Thus, each of the $2^r-1$ hyperplanes of $A$ is the kernel of one of the $n$ coordinate functionals. It follows that $2^r-1\le n$ and hence $r\le b$.

To exhibit a minimal witness of dimension $b$, let $S\in\F_2^{n\times b}$ have every nonzero vector of $\F_2^b$ as a row, exactly once. Its column space has dimension $b$ and full support. For every proper subspace $T<\F_2^b$, some nonzero linear functional vanishes on $T$. That functional appears as a row of $S$, so the corresponding coordinate vanishes throughout the subcode $\inset{Sx:x\in T}$. Every proper subcode of $\operatorname{col}(S)$ therefore fails to have full support. This proves that the locality is exactly $b$.

\paragraph{The probability threshold.}
Let $G\in\F_2^{n\times k}$ have independent uniform entries. Its column space has full support exactly when every row is nonzero, so
$$
\PR{\operatorname{col}(G)\in\cP_b}
=(1-2^{-k})^n\eperiod
$$
For each nonzero $x\in\F_2^k$, we have $\PR{Gx=0}=2^{-n}$. Taking a union bound over $x$ gives $\PR{\rank G<k}<2^{k-n}$. Conditioned on full column rank, $\operatorname{col}(G)$ is a uniformly random $k$-dimensional code. Conditioning changes any event probability by at most the probability of rank deficiency, and hence
$$
\left|\PR{C_k\in\cP_b}-(1-2^{-k})^n\right|
\le2^{k-n}\eperiod
$$
As $b\to\infty$, the unconditioned probabilities at $k=b$ and $k=b+1$ tend to $e^{-1}<1/2$ and $e^{-1/2}>1/2$, respectively. The conditioning errors tend to zero. Monotonicity in $k$ therefore gives $\thresholddim(\cP_b)=b+1$ for all sufficiently large $b$.

\paragraph{The lower bound on the expectation threshold.}
Let $\cA'$ consist of all full-support subspaces of dimensions $1,\ldots,b$. Since every minimal witness has dimension at most $b$, this family is a cover. We will show that its cost tends to zero at
$$
k_-:=b-\lceil\log_2 b\rceil-3\eperiod
$$

Write $N_r$ for the number of full-support $r$-dimensional subspaces. Each has $|\mathrm{GL}_r(\F_2)|$ ordered basis matrices, all with nonzero rows. Counting all matrices with nonzero rows therefore gives
$$
N_r
\le\frac{(2^r-1)^n}{|\mathrm{GL}_r(\F_2)|}
\le4\cdot2^{r(n-r)}e^{-n2^{-r}}\eperiod
$$
For the second inequality, we used $(1-x)^n\le e^{-nx}$ and
$$
|\mathrm{GL}_r(\F_2)|
=2^{r^2}\prod_{j=1}^r(1-2^{-j})
>\tfrac14\,2^{r^2}\eperiod
$$
Indeed, after separating the factor $1/2$, the remaining product is at least $1-\sum_{j=2}^r2^{-j}>1/2$.

For sufficiently large $b$, we have $k_-\ge1$ and $n2^{-k_-}\ge4b$. Consequently,
$$
\covercost_{k_-}(\cA')
\le4\sum_{r=1}^b
e^{(\ln 2)r(k_--r)-n2^{-r}}\eperiod
$$
We bound the terms with $r\le k_-$ and $r>k_-$ separately. In the first case, put $j:=k_--r\ge0$. The exponent satisfies
$$
(\ln 2)r(k_--r)-n2^{-r}
\le(\ln 2)bj-4b2^j
\le-3b\ecomma
$$
using $j\le2^j$. In the second case, put $j:=r-k_-\ge1$ and discard the negative term $-n2^{-r}$. The exponent is then at most $-(\ln 2)(k_-+1)j$. Summing gives
$$
\covercost_{k_-}(\cA')
\le4be^{-3b}
+\frac{4\cdot2^{-(k_-+1)}}{1-2^{-(k_-+1)}}
=o(1)\eperiod
$$
Thus, the cover costs less than $1/2$ at $k_-$ for sufficiently large $b$, and
$$
\expectationdim(\cP_b)
\ge k_-+1
=b-\lceil\log_2 b\rceil-2\eperiod
$$

\paragraph{The upper bound on the expectation threshold.}
We will construct a pseudorandom ensemble consisting entirely of full-support codes and use it to show that every cover has large cost.

Put
$$
d:=b-\lfloor\log_2 b\rfloor,
\qquad
k_+:=d+9\eperiod
$$
For sufficiently large $b$, we have $k_+\le n$. We will construct a random subspace $W$ that always satisfies $\cP_b$ and obeys
\begin{equation}\label{eq:full-support-containment}
\PR{U\le W}\le2^{-(n-k_+)\dim U}
\qquad\text{for every }U\le\F_2^n\eperiod
\end{equation}
To deduce $\expectationdim(\cP_b)\le k_+$ from \cref{eq:full-support-containment}, let $\cA'$ be any cover of the minimal witnesses. Every realization of $W$ contains a minimal witness and therefore a member of $\cA'$. The union bound gives
$$
1
\le\sum_{U\in\cA'}\PR{U\le W}
\le\sum_{U\in\cA'}2^{-(n-k_+)\dim U}
=\covercost_{k_+}(\cA')\eperiod
$$
Thus, every cover has cost at least one at dimension $k_+$.

We will construct $W$ as the column space of a random matrix $M\in\F_2^{n\times d}$. The sampled subspaces will have dimension at most $d$, and we compare their containment probabilities with the RLC upper bounds at dimension $k_+=d+9$. We first sample the rows of $M$ independently and uniformly from $\F_2^d\setminus\{0\}$. Since no row is zero, $\operatorname{col}(M)$ always has full support. We will choose an event $E$ of positive probability and define $W$ using the distribution of $M$ conditioned on $E$. Our task is to choose $E$ so that $W$ also satisfies \cref{eq:full-support-containment}.

To find a suitable conditioning event, we examine the probabilities of equations $MB=N$. Fix a subspace $U$ of dimension $1\le a\le d$, and choose a matrix $N_U$ whose columns form a basis of $U$. The containment $U\le\operatorname{col}(M)$ holds exactly when $MB=N_U$ for some rank-$a$ coefficient matrix $B\in\F_2^{d\times a}$. Indeed, we can choose preimages of the basis vectors as the columns of $B$, and these preimages are linearly independent because their images are. There are at most $2^{da}$ choices of $B$. It therefore suffices to choose $E$ so that
\begin{equation}\label{eq:full-support-conditioned-projection}
\PR{MB=N\mid E}\le2^{-(n-9)a}
\end{equation}
for every $1\le a\le d$, every rank-$a$ coefficient matrix $B$, and every $N\in\F_2^{n\times a}$. Once \cref{eq:full-support-conditioned-projection} holds, the union bound gives
\begin{equation}\label{eq:full-support-containment-from-projections}
\PR{U\le W}
\le2^{da}2^{-(n-9)a}
=2^{-(n-k_+)a}\eperiod
\end{equation}

We first compute $\PR{MB=N}$ before conditioning. For a matrix $N$, let $z(N)$ denote its number of zero rows. Fix a rank-$a$ coefficient matrix $B\in\F_2^{d\times a}$, where $1\le a<d$. Among the $2^d-1$ possible rows of $M$, each nonzero vector of $\F_2^a$ has $2^{d-a}$ preimages under $x\mapsto xB$, whereas zero has $2^{d-a}-1$. Multiplying the probabilities for the independent rows gives
\begin{equation}\label{eq:full-support-projection-probability}
\PR{MB=N}
=2^{-na}(1-2^{-d})^{-n}(1-2^{a-d})^{z(N)}\eperiod
\end{equation}
For $a=d$, the probability is zero if $N$ has a zero row, and otherwise equals $2^{-nd}(1-2^{-d})^{-n}$.

The factor $2^{-na}$ in \cref{eq:full-support-projection-probability} is the probability of $N$ under the uniform distribution on $\F_2^{n\times a}$. To obtain \cref{eq:full-support-conditioned-projection}, we need to limit how much the remaining factors increase this probability. Put
$$
\lambda:=n2^{-d}
=(1-2^{-b})2^{\lfloor\log_2 b\rfloor}\eperiod
$$
For sufficiently large $b$, we have $d/3\le\lambda\le2d$, so $(1-2^{-d})^{-n}=2^{O(d)}$. When $a\ge d/2$, this is already a factor of $2^{O(a)}$.

For $a<d/2$, we use the zero rows of $N$ to compensate for this increase. To motivate the required number of zero rows, consider the natural logarithm of the correction factor in \cref{eq:full-support-projection-probability}. Ignoring lower-order terms for the moment, it is
$$
\lambda-2^{a-d}z(N)\eperiod
$$
The two terms cancel when $z(N)=n2^{-a}$. If the number of zero rows falls below this value by a fraction $a/d$, the remaining contribution is approximately $\lambda a/d=O(a)$. This suggests requiring
\begin{equation}\label{eq:full-support-zero-row-condition}
z(N)\ge\inparen{1-\frac ad}n2^{-a}\eperiod
\end{equation}
We therefore let $E$ be the event that \cref{eq:full-support-zero-row-condition} holds with $N=MB$ for every rank-$a$ coefficient matrix $B$ with $1\le a<d/2$. We first show that $\PR{E}=1-o(1)$, then verify the numerical bound in \cref{eq:full-support-conditioned-projection}.

Fix $a<d/2$ and a rank-$a$ coefficient matrix $B$. The variable $z(MB)$ is binomial with mean
$$
\mu=n2^{-a}\frac{1-2^{a-d}}{1-2^{-d}}\eperiod
$$
Write $\mu_0:=n2^{-a}$. For sufficiently large $d$, we have $2^{a-d}\le a/(2d)$ uniformly over $1\le a<d/2$. Hence
$$
\mu\ge\inparen{1-\frac{a}{2d}}\mu_0,
\qquad
\inparen{1-\frac ad}\mu_0
\le\inparen{1-\frac{a}{2d}}\mu\eperiod
$$
Thus, violating \cref{eq:full-support-zero-row-condition} requires a downward deviation of at least an $a/(2d)$ fraction of the mean. The Chernoff bound gives
\begin{align*}
\PR{z(MB)<(1-a/d)\mu_0}
&\le\exp{-\Omega\inparen{\frac{a^2}{d^2}n2^{-a}}}\\
&\le\exp{-\Omega\inparen{\frac{2^{d/2}}d}}\eperiod
\end{align*}
For the second inequality, we use $n2^{-a}=\lambda2^{d-a}\ge(d/3)2^{d/2}$ and $a\ge1$. The implicit constants are absolute and uniform in $a$ and $B$. There are at most $2^{O(d^2)}$ coefficient matrices across these dimensions, so a union bound gives
$$
\PR{E^c}
\le\exp{O(d^2)-\Omega\inparen{\frac{2^{d/2}}d}}
=o(1)\eperiod
$$
In particular, for sufficiently large $b$,
\begin{equation}\label{eq:full-support-good-event-probability}
\PR{E}\ge\tfrac12\eperiod
\end{equation}

We now prove \cref{eq:full-support-conditioned-projection}. By \cref{eq:full-support-good-event-probability}, conditioning increases probabilities by at most two, so we first bound $\PR{MB=N}$ for outcomes compatible with $E$. Fix a rank-$a$ coefficient matrix $B$ and a matrix $N\in\F_2^{n\times a}$.

Suppose first that $a<d/2$. If $N$ fails \cref{eq:full-support-zero-row-condition}, then $MB=N$ is incompatible with $E$, so its conditional probability is zero. Otherwise, substituting \cref{eq:full-support-zero-row-condition} into \cref{eq:full-support-projection-probability} and using $-\ln(1-x)\le x/(1-x)$ and $\ln(1-x)\le-x$ gives
\begin{align*}
\log_2\inparen{\frac{\PR{MB=N}}{2^{-na}}}
&\le\frac{\lambda}{\ln 2}
\inparen{\frac1{1-2^{-d}}-\inparen{1-\frac ad}}\\
&=\frac{\lambda}{\ln 2}
\inparen{\frac{2^{-d}}{1-2^{-d}}+\frac ad}\\
&\le1+\frac{2a}{\ln 2}
<4a\eperiod
\end{align*}
Here we used $\lambda\le2d$ and took $d$ sufficiently large that $2d\,2^{-d}/((1-2^{-d})\ln 2)\le1$. Thus, for every $N$ satisfying \cref{eq:full-support-zero-row-condition},
\begin{equation}\label{eq:full-support-small-projection}
\PR{MB=N}\le2^{-na+4a}\eperiod
\end{equation}

For $d/2\le a\le d$, we instead use
\begin{equation}\label{eq:full-support-large-projection}
\PR{MB=N}
\le2^{-na}(1-2^{-d})^{-n}
\le2^{-na+4d}
\le2^{-na+8a}\eperiod
\end{equation}
For $a<d$, the first inequality follows by bounding the last factor in \cref{eq:full-support-projection-probability} by one. For $a=d$, it follows because multiplication by $B$ permutes the nonzero row vectors, so each prescribed row has probability at most $1/(2^d-1)$. The second inequality follows from
$$
\log_2(1-2^{-d})^{-n}
\le\frac{\lambda}{(1-2^{-d})\ln 2}
\le4d\eperiod
$$
Multiplying the bounds in \cref{eq:full-support-small-projection,eq:full-support-large-projection} by at most two to account for conditioning, and using $2\le2^a$, proves \cref{eq:full-support-conditioned-projection}.

We can now define $W$ using the distribution of $\operatorname{col}(M)$ conditioned on $E$. Every realization of $W$ has full support. By \cref{eq:full-support-containment-from-projections}, it also satisfies \cref{eq:full-support-containment} for $1\le\dim U\le d$. For $\dim U>d$, containment has probability zero, and for $U=\{0\}$ the bound holds with equality. Hence
$$
\expectationdim(\cP_b)
\le k_+
=b-\lfloor\log_2 b\rfloor+9\eperiod
$$
Together with the lower bound and the probability-threshold calculation, this completes the proof.
\end{proof}

\end{document}